\documentclass{article}
\usepackage[margin=1in]{geometry}
\usepackage[utf8]{inputenc}
\usepackage{xcolor}
\usepackage{graphicx}
\usepackage{float}
\usepackage{wrapfig}
\usepackage[numbers, sort&compress]{natbib} % bibliography
\usepackage{url} % bibliography
\usepackage{amsmath,amssymb,amsfonts,amsthm} % Related to math
\usepackage{multirow} % tables
\usepackage{ifthen} % nice if else statements
\usepackage{BOONDOX-calo}
\usepackage{subcaption} % Add this in your preamble
\usepackage{booktabs} % for nice tables
\usepackage{array}
\usepackage[table, dvipsnames]{xcolor}   % color support for tables
\usepackage{colortbl} 
\usepackage{enumitem}

\renewcommand{\arraystretch}{1.3} % etwas mehr Zeilenhöhe
\definecolor{smcolor}{HTML}{fc9f00}
\definecolor{ringcolor}{HTML}{36e236}
\definecolor{densecolor}{HTML}{7304f9}
\definecolor{mydarkred}{RGB}{217,29,0} % dark red

\title{Mass Manipulation in Simulated Social Networks:\\Dominating vs. Diversifying Attention}

\usepackage{authblk} % authors handling
\author[1,2,*]{Viktoria Kainz}
\author[3]{Justin Sulik}
\author[1,2]{Anna Neudert}
\author[1,2,4]{Torsten Enßlin}
    
\affil[1]{Max Planck Institute for Astrophysics, Garching}
\affil[2]{Faculty of Physics, Ludwig Maximilian University of Munich}
\affil[3]{Cognition, Values \& Behavior Lab, Munich Interactive Intelligence Initiative, LMU Munich}
\affil[4]{German Center for Astrophysics, Görlitz}
\affil[*]{Corresponding Author: \texttt{viktoria@mpa-garching.mpg.de}}

\begin{document}

\maketitle

\begin{abstract}

Modern information environments, especially social media, are highly complex systems that exceed individual processing capacities such as humans' limited attention. This environment/cognition mismatch can increase susceptibility to misinformation, which various actors exploit for anti-social (including anti-democratic or anti-science) aims. This raises the question of how to feasibly sustain societal resilience against misinformation, though the challenge is to find strategies that respect individuals' cognitive limitations. We investigate whether a simple behavioral rule---topic diversification---can enhance collective performance and mitigate vulnerability. In an agent-based model that includes a deceptive mass-influencing agent (MIA), we compare two attention-distribution strategies: (A) acquaintance-based topic selection, where agents return to familiar content, and (B) randomized topics, which diversify attention. We also track dynamics across different network structures. We find that under acquaintance-based topics, a central MIA advances its propaganda effectively, causing volatile and polarized opinions through repeated exposure and echo chambers. Under randomized topics, this leverage disappears: the MIA's influence collapses across all network structures, and opinions become stable and broadly aligned with reality. These results, while deriving from simple simulations, align with realistic theories of bounded rationality and collective cognition, further suggesting a cognitively feasible, easy-to-monitor and robust strategy: distribute attention to combat misinformation.
\end{abstract}

\paragraph*{Keywords}
Misinformation, Attention, Communication Networks, Collective Cognition, Social Learning, Polarization, Echo-Chambers, Bayesian Reasoning, Agent-Based Model

\section{Introduction}

When forming opinions, the information people rely on---and the sources we trust---play a decisive role in shaping our beliefs \cite{boumans2025fostering}. However, decisions about information and its sources are increasingly complex: recent decades have seen a shift from a low-choice media landscape with limited availability of information sources (for the majority, restricted to public broadcasting or a few newspaper publishers) to a high-choice landscape with an oversupply of information, from which every person now has to select individually. This situation has been described as ``epistemic flooding'' \cite{anderau2023fake}.

Consequences of this environment include knowledge inequalities, fragmentation and polarization, and declining news quality \cite{van-aelst2017political, bentivegna2020rethinking, karlsen2020high, stromback2022low, panek2016high}. Moreover, various actors can exploit this new media landscape for their own purposes, where ``actors'' may be individuals such as influencers or politicians, or orchestrated campaigns including artificial intelligence-driven bots \cite{woolley2022digital, muth2023social, shmalenko2021impact, marquart2020following, khaund2021social, keller2019social}. 

In this information-rich and possibly deceptive environment, people now face two key difficulties when trying to build informed opinions: 1) psychological biases when learning from social information; and 2) limited attention. 

Biases, which include a preference for belief-consistent over contradicting information, a tendency to overweight negative information or a need for social acceptance, are well-known drivers for many of the aforementioned effects such as fragmentation and polarization, and biases are often amplified by feed algorithms on social media \cite{munro1997biased, jagiello2018bad, barkoczi2016social, robertson2024inside}. 

However, even if the public were able to overcome these and other cognitive biases, basic processing constraints such as limited attention make it impossible to adequately process the sheer volume of information. For instance, logical contradictions are difficult to identify in a flood of information, which opens the path for parallel, potentially contradicting ``truths'' or world-views to coexist \cite{hameleers2025entering, abdelzaher2020paradox, qiu2017limited}. Opinion landscapes become scattered and individuals often remain within small, like-minded bubbles or echo chambers \cite{nguyen2020echo}.

The combination of fragmented attention, selective trust and cognitive biases creates an environment in which misinformation can spread and persist, even in the presence of clear contradictory evidence \cite{lazer18, del2016spreading}.  We are thus confronted with what has been described as ``post-truth era'' \cite{lewandowsky2017beyond, yerlikaya2020social, sawyer2018post}, a situation in which truth becomes hard to identify reliably\footnote{This need not imply a stronger sense of ``post-truth'' in which individuals cease to care about truth at all. Our models assume rational agents that try to form true beliefs.}.

Yet, for any actor aiming to achieve large-scale influence---whether in politics, activism, social movements or any other field---reaching and convincing a broad majority is essential. This raises two complementary questions: how do misinformation strategies work by exploiting limited attention, and how can individuals or societies resist such exploitation?

\subsection{Modeling Deception in an Information-Rich World}

In this study, we build upon an agent-based model, the ``Reputation Game Simulation''. The relevance of this model to the current issue is that the agents vary in their baseline honesties (or propensities to share truthful information when interacting), so the central challenge is for agents to infer just how reliable others are. The dynamic flow of information means that even rational agents can mistake reliable sources as liars, or liars as reliable. These mistakes are sometimes simply caused by deliberate deception, yet often reflect more nuanced and complex causes, given how diverse communication strategies may accidentally align so as to make some piece of shared information seem more or less truthful (and its source more or less reliable, as a knock-on effect). 

With this foundation, the model has already demonstrated the ability to reproduce real-world phenomena such as echo chambers, polarization or freezing of group opinions when individuals were exposed to malicious, misinformation-spreading actors in their immediate neighborhood \cite{ensslin22}. Further, the model has also produced illusory-truth effects or cluster formation when scaled up to larger networks, where it is difficult for individuals to monitor information flow across the whole network \cite{kainz2022information}. 

The current study extends the previous model with the introduction of deliberate misinformation campaigns that have the potential to impact the entire network, in order to understand the mechanisms behind manipulative strategies (as explored in \cite{ensslin22}) on network-level opinion formation processes (following \cite{kainz2022information}). By understanding these mechanisms, we aim to identify strategies for resisting them. In particular, the approach includes two features common in misinformation campaigns \cite{paul2016russian, woolley2018computational, campos2019cambridge, seeme2025ignorance}: persistent propaganda (an immense volume of aligned information, often achieved by bots in the real world) and targeted manipulation (micro-targeting of individuals with personalized content). 

Finally, as the issue hinges on the flow of social information, different network structures likely impact results. We focus on small-world topologies as they have been widely documented in human social interactions \cite{de1978contacts, schnettler2009structured}, popularized by Milgram's ``six degrees of separation'' experiment in the mid-20th century \cite{milgram1967small} but still found in social media platforms today \cite{mislove2007measurement, dimitri2023facebook, kwak2010twitter}. 

We contrast these with ring and dense networks \cite{zollman07, zollman10}. These represent theoretical extremes of minimal and maximal connectivity (on the assumption that all agents are connected to at least one other agent), allowing us to disentangle which effects are specific to certain network topologies and which are general features of social communication under uncertainty. 

\subsection{Resistance and Self-Stabilization}\label{subsec-resistance}

Understanding how misinformation strategies work is just one part of the story; another, even more important part is understanding how individuals and society as a whole should behave and remain resilient in this  environment. Many common proposals can be grouped into two broad approaches: 1) structural/algorithmic changes; and 2) improving individual critical thinking. 

The first group suggests ways to modify information flow or its manner of presentation, such as by adjusting feed algorithms on social media, enriching displayed content with accuracy tags, or similar platform-driven actions \cite{burel2024exploring, kozyreva2024toolbox, fernandez2024analysing}. However, since these proposed remedies rely on platform providers, their usefulness is limited when providers have conflicts of interest. 

The second group therefore focuses on actions that individuals can take, such as increasing their media literacy skills, engaging in reflective thinking to mitigate cognitive biases, or trying to understand views that conflict with their own \cite{french2025impact, alon2024fighting, machete2020use, pennycook2020fighting}. However, although in principle achievable by individuals regardless of media platforms' conflicts of interests, these strategies are often difficult to apply in everyday life as they require significant effort and awareness whenever facing new information.

We propose a third alternative, still within individual control but based on a simple rule: \emph{diversify the topics you engage with}. This might be easier said than done as people often follow fixed routines, regularly consuming the same information channels which may include focusing on a limited set of recurring topics. Still, we believe that this strategy may be simpler than the alternatives, such as identifying the single ``best'' source or exhaustively comparing numerous sources, and constantly engaging critical or reflective reasoning while avoiding cognitive bias. One does not need to engage critical reasoning to diversify one's range of topics, and any psychological factors that may hamper diversification of topics (such as habits while consuming media) also apply to the usual alternatives, for instance if one habitually consumes media without engaging critical reflection. 

In this study, we evaluate the collective outcomes of this proposal in a simulated society, leaving empirical testing of its psychological feasibility for future work. Even so, our approach comes with two key advantages: 1) it supports an exploratory, curiosity-driven mindset that encourages people to search for new information instead of employing an instructional tone warning of risks; and 2) non-compliance is easy to detect: if your news feed contains barely more than a single topic, some diversification might be advisable. This makes it an eminently practical starting point, applicable even with low mental capacity and minimal time effort.

We therefore analyze two conditions, focusing on how susceptibility to mass misinformation campaigns changes when agents either (A) select topics based on previous acquaintance, thus focusing on familiar topics they regularly encounter; or (B) diversify the topics they engage with. Finally, we also investigate the transition between both conditions by varying the faction of agents using either A or B, to show not only how susceptibility changes for the whole network but also for the individual agents using each strategy.

\section{Model} \label{sec:model}

This study extends the Reputation Game Simulation \cite{ensslin22, ensslin2023simulating, kainz2022information}\footnote{For readers who wish to know how this relates to earlier work: we use the exact same setup as in \cite{kainz2022information} with the ``friendship affinity'' parameter $F$ set to $0$. Parameter studies and justification of certain assumptions can be found in \cite{ensslin22} and \cite{ensslin2023simulating}.}. We will describe the basic mechanics of the general model in \S\ref{subsec:the-reputation-game}, and study specific settings in \S\ref{subsec:experiment-specific-setup}. For other details of the model not directly relevant to the current topic, readers should consult the references provided above. 

\subsection{The Reputation Game Simulation}\label{subsec:the-reputation-game}

The Reputation Game Simulation (RGS) is an agent-based model simulating communication dynamics. We consider a fixed set of agents $\mathcal{A}$, here a population $|\mathcal{A}|=50$ individuals. Agents' communicative interactions occur over several rounds. In each round, every agent initiates exactly one bout of communication. The order in which agents do so is randomized per round. 

When initializing a communicative interaction, an agent $a \in \mathcal{A}$ chooses a set of agents $\mathcal{b} \subseteq \mathcal{A}\setminus \{a\}$ to talk to, as well as an agent $c \in \mathcal{A}$ as conversation topic: agents $a$ and $\mathcal{b}$ gossip about agent $c$, and this includes agents  talking about themselves or about their interaction partners. Talking about topic $c$ just means reporting how many times an agent believes agent $c$ to have told the truth or to have lied previously, though this report need not reflect their actual opinion (further discussed under `Communication' below). As the point of this model is to understand how reputations for honesty or lying evolve over collective interactions, having the topics just be agents themselves simply focuses all communication on those reputations. 

\begin{figure}[H]
\centering
\includegraphics[width=0.9\textwidth]{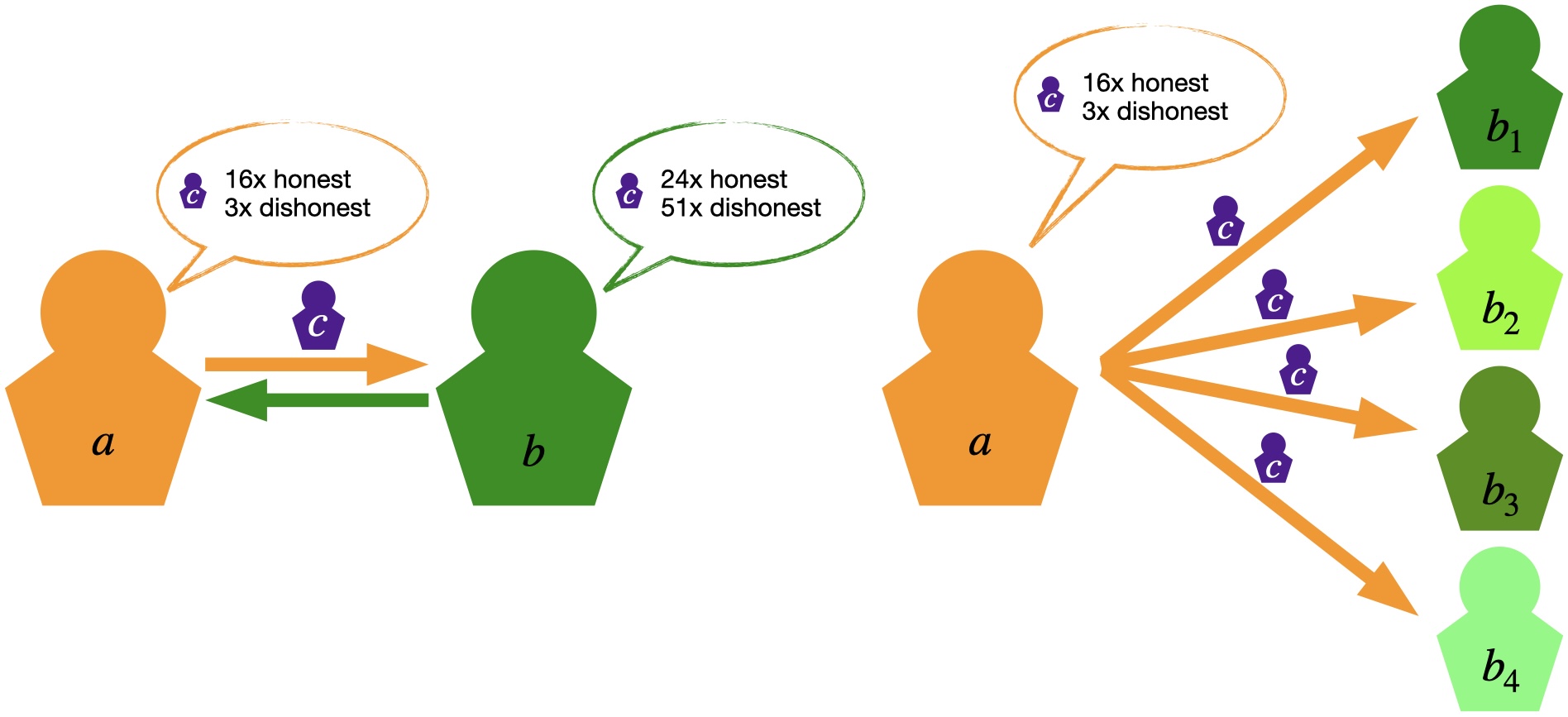}
\caption{Communication structure in the RGS with initiator $a$, receiver(s) $b$ and topic $c$, as well as exemplary statements about $c$'s reputation. Left: one-to-one interaction, Right: one-to-many interaction.}
\label{fig:agent_conversations}
\end{figure}

In this model, there are two types of interactions: one-to-one and one-to-many, both illustrated in figure \ref{fig:agent_conversations}. We simulate a roughly equal number of each type, randomizing this choice prior to each interaction. In one-to-one interaction there is only one recipient $\mathcal{b} = \{b\}$ to whom conversation initiator $a$ transmits its message about $c$. Subsequently, $b$ answers about the same topic to $a$. In one-to-many interaction, there are multiple receivers $\mathcal{b} = \{b_1, b_2, ...\}$ that hear $a$'s message. However, they do not respond, as this is more of a public announcement or broadcast than a conversation. 

After an interaction, all agents that received some information (i.e., $a$ and $b$ in one-to-one conversations or all recipients $\mathcal{b}$ in one-to-many interactions) update their knowledge accordingly. How they do so is explained under `Updating knowledge' below. The game continues like this for a fixed number of rounds as agents' opinions of each other evolve.

\paragraph{Communication: choosing partners and topics}

Agents choose both communication partners and topics based on acquaintance, causing them to form and adhere to social subgroups within the larger network. Acquaintance is therefore measured by the number of times they have interacted with an agent as well as the number of messages they have received about it, leading to an acquaintance of agent $i$ with agent $j$ defined as 

\begin{align}
\mathrm{Acq}_{ij} = n^\mathrm{c}_{ij} + n^\mathrm{m}_{ij},
\end{align}

\noindent where $n^c_{ij}$ is the number of times agent $j$ communicated to $i$ and $n^m_{ij}$ is the number of times $i$ has received a message about $j$. The probabilities for choosing $b$ as communication partner or $c$ as communication topic are then given by 

\begin{align}
\label{eq:draw_b}
P(\text{$a$ chooses $b$ as partner}|\mathrm{Acq}_{ab},S_a) &\propto \mathrm{Acq}_{ab}^{S_a} \text{\ \ \ and} \\
\label{eq:draw_c}
P(\text{$a$ chooses $c$ as topic}|\mathrm{Acq}_{ac}, S_a) &\propto \mathrm{Acq}_{ac}^{S_a},
\end{align}

\noindent with $S_i\geq0$ being an intrinsic property of agent $i$ called ``shyness'', fixed throughout the simulation. This parameter controls how much weight an agent puts on acquaintance. Agents prefer to talk to/about agents they know well, and the higher their shyness, the stronger this tendency. Additionally, shyness defines how many receivers an agent targets in one-to-many interactions, as each time, the number of recipients is drawn from the distribution
\begin{align}
P(\text{number of recipients} = N_b|S_a) \propto N_b^{-S_a}.
\end{align}

With a default value of $S_a=10^{0.5}$ for all ordinary agents, the typical number of recipients is $\langle N_b \rangle \approx 2.93$, contributing to a reasonable typical size of social subgroups around $4$ agents ($2.93$ rounds to $3$, plus the speaker) in a network of $50$. 

\paragraph{Communication: message contents and tells}

Communication here is about opinion exchange. In the model, this can be intentional, via sending explicit messages. It can also be unintentional, analogous to ``tells'' in a game of poker, as when someone accidentally signals that they are bluffing by scratching their ear or blushing. It seems implausible that every individual in a complex informational ecosystem should always and unfailingly be in full, intentional control of every piece of information they reveal, so these tells simply allow for leakage via a second channel of information. 

Following the basic RGS \cite{ensslin22}, agents only communicate about each others' (and their own) honesty $x$. Every agent $i$ has an intrinsic honesty $x_i \in [0,1]$ which is fixed throughout the simulation. This defines how often an agent shares the truth, meaning that it lies in a proportion $1-x_i$ of its messages. Specifically, if a random draw from binomial $B(1, x_i)$ yields 1, the agent shares an honest statement. If so, agent $a$ communicates its current opinion about topic $c$---how honest $a$ currently thinks $c$ is---to receiver(s) $\mathcal{b}$. Bearing in mind that agents' sincere opinions may not accurately represent the true state of affairs (after all, they are somewhat uncertain about others' reputations), honesty does not necessarily imply accuracy. 

In case of a dishonest statement, the speaker is free to lie, misrepresenting their actual opinion. Plausibly, when lying, they would distort messages in their favor. To see what this means, we must first unpack the technical details of the messages (and the opinions they are meant to reflect), as well as agents' goals in sharing their opinions.  

Opinions and messages both have the form of a beta distribution defined by $\mu, \lambda \in (-1, \infty)$ that can be understood as the number of honest ($\mu$) or dishonest statements ($\lambda$) an agent is believed to have made. An opinion about someone's honesty can thus be modeled as a probability distribution

\begin{align}
P(x|I) = \mathrm{Beta}(x|\mu, \lambda) \propto x^\mu (1-x)^\lambda,
\end{align}

\noindent where $I=(\mu, \lambda)$ encodes the information an agent has (or pretends to have) about someone else. The beta distribution has mean $\overline{x} = \frac{\mu+1}{\mu + \lambda + 2}$, which we call the ``reputation'' of an agent in the eyes of another agent holding this opinion. All agents aim to maximize their own reputations. 

By deploying strategic lies, any agent with less than 100\% honesty can deliberately boost its reputation by over-stating its own honesty, i.e. increasing $\mu$ in a message where it is the topic. An agent can also boost its own reputation indirectly by similarly over-stating the honesty of other agents that it believes might share a good report of it (or conversely, under-stating the reputation of agents that it thinks might share a bad report of it by increasing $\lambda$).

At the same time, an agent should not exaggerate too much when lying, lest its distortions get detected. To handle this, we implement a minimal Theory of Mind, whereby a lying agent will try to estimate what its receiver(s)\footnote{In case of several receivers, the agents use the average of all receivers' estimated opinions, see equation $6$ in \cite{kainz2022information}.} likely believe about topic $c$ to make it sound plausible, and only deviating a little from this estimated anchor\footnote{Mathematical definitions of how agents update their Theory of Mind or decide how large the deviation can be, are given in appendix B2 in \cite{ensslin22}. In short: they track everyone's opinions from previous interactions and adapt lie sizes to the typical opinion variability currently surrounding them. The latter might change over time and is measured by the median surprise an agent experiences when receiving information from others.}. Thus, agents communicate explicitly by sharing a message in the form $(\mu, \lambda)$, which is either their honest opinion or a strategically constructed lie. 

The second information channel, the unintentional tells, are comparatively more straightforward. Whenever an agent lies, there is a $10\%$ chance that it will give off an additional signal (and $0\%$ chance of this happening when honest), much like a blush. This just provides others with reliable (yet stochastic) information that they can use in trying to learn who among them is being honest. 

\paragraph{Updating Knowledge}\label{par: updating-knowledge}

The first task when receiver $b$ seeks to update its beliefs, in light of a newly communicated message from initiator $a$, is for $b$ to estimate the credibility of $a$'s message $P(\text{h}|d)$: the probability of the message having been honest $\text{h}$, given the available data $d$ (message content, the presence/absence of a tell and any current knowledge about the speaker and topic). This is done in a Bayesian manner, using the receiver's estimate of the speaker's general honesty $\overline{x}_{ba}$, the surprise of the message $\mathcal{S}$, the tell if present and whether the statement was a confession\footnote{A confession is defined as a statement about the speaker itself, presenting a worse reputation than what the receiver already assumes. In this case, the message should be believed regardless of other factors, since even if it was a lie, believing it brings the receiver closer to the truth. Further, the message's surprise is defined as the normalized distance between current knowledge $I$ about the topic and incoming message $J$, proportional to $\mathrm{KL}(J|I)$, with $\mathrm{KL}$ being the Kullback-Leibler Divergence \cite{kullback1951}. A derivation of the formula can be found in \cite{ensslin22}, appendix B1.}:

\begin{align}
\label{eq:credibility}
P(\text{h}|d) 
 &= \frac{1}{1 + R\,(\overline{x}_{ba}^{-1} - 1)}
  \qquad\text{with } 
  R = 
  \begin{cases}
  0.9 \mathcal{S}^2 / 2, & \text{no tell \& no confession} \\
  \infty , & \text{tell \& no confession} \\
  0, & \text{confession.}
  \end{cases}
\end{align}

Having made this credibility assessment, the second step is for agent $b$ to update its beliefs accordingly, regarding both the speaker and the topic ($I_{ba}$ and $I_{bc}$, respectively). For both updates we have to consider the  possibilities of an honest statement or a lie, weighting these according to aforementioned credibility $P(\text{h}|d)$:

\begin{align}
\label{eq: update_topic}
P(x_{c}|I^{\text{updated}}_{bc}) &= P(\text{h}|d) \mathrm{Beta}(x_c|\mu_{bc}+\Delta \mu, \lambda_{bc}+\Delta \lambda) + (1-P(\text{h}|d)) \mathrm{Beta}(x_c|\mu_{bc}, \lambda_{bc})\\
\label{eq: update_speaker}
P(x_{a}|I^{\text{updated}}_{ba}) &= P(\text{h}|d) \mathrm{Beta}(x_a|\mu_{ba}+1, \lambda_{ba}) + (1-P(\text{h}|d)) \mathrm{Beta}(x_a|\mu_{ba}, \lambda_{ba}+1).
\end{align}

Here, indices always mark that it is agent $b$'s estimation of $a$ or $c$'s honesty measures. $\Delta \mu$, $\Delta \lambda$ in the first row and the $+1$ in the second row are the additional information that could be learned from interactions, while $(\mu_{ba},\lambda_{ba})$ or $(\mu_{bc},\lambda_{bc})$ represent the receiver's prior knowledge. Lastly, we compress these bimodal, updated belief states into the form of a single beta distribution both for computational reasons and to mimic the variety of mental shortcuts that humans naturally use \cite{Tversky&Kahneman}. We do so choosing the optimal approximation, such that as little information is lost as possible\footnote{Specifically, this is done by minimizing the Kullback-Leibler Divergence between the approximated, simpler beta distribution and the correct posterior in equations \ref{eq: update_topic} and \ref{eq: update_speaker}. Computational details can be found in Appendix A4 in \cite{ensslin22}.}.

\subsection{Experiment Specific Setup}
\label{subsec:experiment-specific-setup}

In this study we want to investigate the consequences of deliberate misinformation campaigns: the effects of propaganda and the resilience of different network types to that propaganda, focusing here on the specific resilience strategy---diversifying communication topics---described in the Introduction (\S\ref{subsec-resistance}).

\paragraph{Modeling Propaganda -- The Mass Influencing Agent (MIA)}
Propaganda is modeled as coming from one agent, the ``Mass Influencing Agent'' (MIA), which tries to convince the whole network that it has a very high reputation, despite being $0\%$ honest in reality. The challenge for the MIA is to establish a belief in the network that contradicts the frequent, exposing tells. The MIA-strategy implemented here is a combination of two strategies that proved to be efficient for the same purpose in smaller groups (``manipulative'' and ``dominant'' strategies in \cite{ensslin22, ensslin2023simulating}), combined with a preference for larger audiences. This results in different probability distributions for choosing interaction partners and topics, compared to ordinary agents whose defaults are described above: 

\begin{align}
\label{eq:draw_b_mia}
P_\text{MIA}(\text{$a$ chooses $b$ as interaction partner}|I_a) &\propto 
	\begin{cases}
  1-\overline{x}_{ab}, & \text{One-to-One} \\
  \overline{x}_{ab} , & \text{One-to-Many}
  \end{cases}\\
\label{eq:draw_c_mia}
P_\text{MIA}(\text{$a$ chooses $c$ as interaction topic}|\text{$b$ is the partner}) &=
	\begin{cases}
  \delta_{bc}, & \text{One-to-One} \\
  \delta_{ac} , & \text{One-to-Many}
  \end{cases},
\end{align} 

\noindent where $\delta_{ij}$ is the Kronecka-Delta yielding $1$ if $i=j$ and $0$ otherwise. In one-to-many interactions, MIA thus looks for highly reputed agents and promotes itself with positive lies---a very straight-forward propaganda technique. In one-to-one conversations, however, the MIA uses a more subtle strategy, looking for dishonest supporters and flattering them. This way, those agents regard the MIA as someone making positive statements about them, hence will return the favor and become powerful (because dishonest) amplifiers. Note that in order to give those strategic decisions more weight, acquaintance does not influence its choices at all. In addition, the MIA also has a reduced shyness of $S=10^{0.3}$, so in one-to-many interactions it talks to approximately $5$ to $6$ agents on average, giving it a slightly larger audience than ordinary agents.

The effect of all individual components (strategic decisions, low shyness and switched-off acquaintance) is discussed in the appendix \ref{sec: parameter-study}.

\paragraph{Network Structures}
Initially, the network has no connections. Thus, the network structures emerge from the dynamics of agents' interactions, rather than being hard-coded by us at the start of a simulation run. The default strategies (\S\ref{subsec:the-reputation-game}) result in a small-world network structure. In order to investigate other structures (ring and dense networks%\footnote{A fourth common network structure would be a star network, but this is characterized by the one, central node and can thus not be achieved by specifying all ordinary agents in a uniform way. However, it will turn out, that a MIA in a small-world (or ring) network happens to become this central node, turning them into star shaped networks naturally.}
), we either increase ordinary agents' shyness (yielding a ring network) or switch off the acquaintance mechanism for choosing interaction partners (yielding a dense network). For the former, the higher shyness value is $S=10^{1}$ (instead of the default $S=10^{0.5}$), which leads to barely more than $2$ partners in one-to-many interactions. For the latter, switching off the acquaintance mechanism means that the probability distribution for choosing a certain partner (cf.\ equation \ref{eq:draw_b}) is flat. All three configurations are summarized in table \ref{tab:network_structures}.

\begin{table}[H]
\centering
\renewcommand{\arraystretch}{1.25}

\begin{tabular}{|
  >{\raggedright\arraybackslash}m{3cm}
  %!{\vrule width 1.2pt}
  | >{\centering\arraybackslash}m{5cm}
  | >{\centering\arraybackslash}m{3cm} |
}
\hline
\rowcolor{black!3}
\textcolor{black}{\textbf{Network type}} & \textbf{\textcolor{black}{Acquaintance mechanism for conversation partner}} & \textbf{\textcolor{black}{Shyness}} \\
\hline

\rowcolor{smcolor!40}
\textbf{Small World} & On & Default \\
\hline

\rowcolor{ringcolor!40}
\textbf{Ring} & On & High \\
\hline

\rowcolor{densecolor!40}
\textbf{Dense} & Off & Default \\
\hline
\end{tabular}

\caption{Parameter settings for ordinary agents to achieve different network types. The colors mark the network type both here and in subsequent figures \ref{fig:rep-hist-AcqC}, \ref{fig:degree-rep-AcqC}, \ref{fig:polarization-AcqC}, \ref{fig:rep-hist-noAcqC}, \ref{fig:degree-rep-noAcqC} and \ref{fig:polarization-noAcqC}.
}
\label{tab:network_structures}
\end{table}

\paragraph{Experimental Design}
In this study, we examine the performance of one Mass Influencing Agent (MIA) among $49$ ordinary agents. The agents' intrinsic honesties are linearly assigned from $0\%$ (agent 0) to $100\%$ (agent 49), where the MIA replaces the most dishonest agent $0$, giving it maximal scope to influence others. 

We compare the MIA's performance across three network types (small world, ring, dense) in two conditions (A and B), as well as a transition from A to B, to test whether randomizing topic selection increases the networks' and individuals' resilience to propaganda. The two conditions can be defines as

\begin{enumerate}[label=\Alph*.]
 \item \textbf{Propaganda with Acquaintance-Based Topic Selection:} Ordinary agents act exactly as described in \S\ref{subsec:the-reputation-game}
 \item \textbf{Propaganda with Randomized Topic Selection:} Ordinary agents choose their conversation topic randomly, i.e., flattening the distribution in equation \ref{eq:draw_c}.
\end{enumerate}

To ensure robustness, we run $100$ simulations for each combination of experimental condition and network type, each time with a different random seed. Each simulation runs for $300$ rounds until dynamics are steady. At the start of each simulation, agents have completely uninformative opinions ($I=(0,0)$) and no connections to others. 
\section{Results}\label{sec:results}

First, we present some qualitative descriptions of emergent network structures, both with and without a Mass Influencing Agent (MIA), to illustrate the structural changes the MIA causes to the network (\S\ref{subsec:emergent-network-structures}). Second, we quantitatively investigate the MIA's performance in different networks structures, also exploring causal effects and underlying mechanisms (\S\ref{subsec:reputation-dynamics}). Third, we show that randomizing the conversation topics increases the ordinary agents' resilience against manipulation drastically across all network types (\S\ref{subsec:effects-of-randomized-topic-choice}). Finally, we see that, especially in small world networks, this scales with the fraction of agents randomizing their topics, while paying off immediately for the individuals doing so (\S\ref{subsec: transition-from-A-to-B}). 

For the quantitative analyses in \S\ref{subsec:reputation-dynamics}--\ref{subsec: transition-from-A-to-B} we use a number of metrics---reputation of the MIA, degree of the MIA, opinion volatility, and opinion polarization---that are mathematically defined in Appendix \ref{sec:metrics}.

\subsection{Emergent Network Structures}\label{subsec:emergent-network-structures}

\begin{figure}[H]
\centering
\begin{subfigure}{0.32\textwidth}
  \includegraphics[width=\linewidth, trim=80 50 80 50,
    clip]{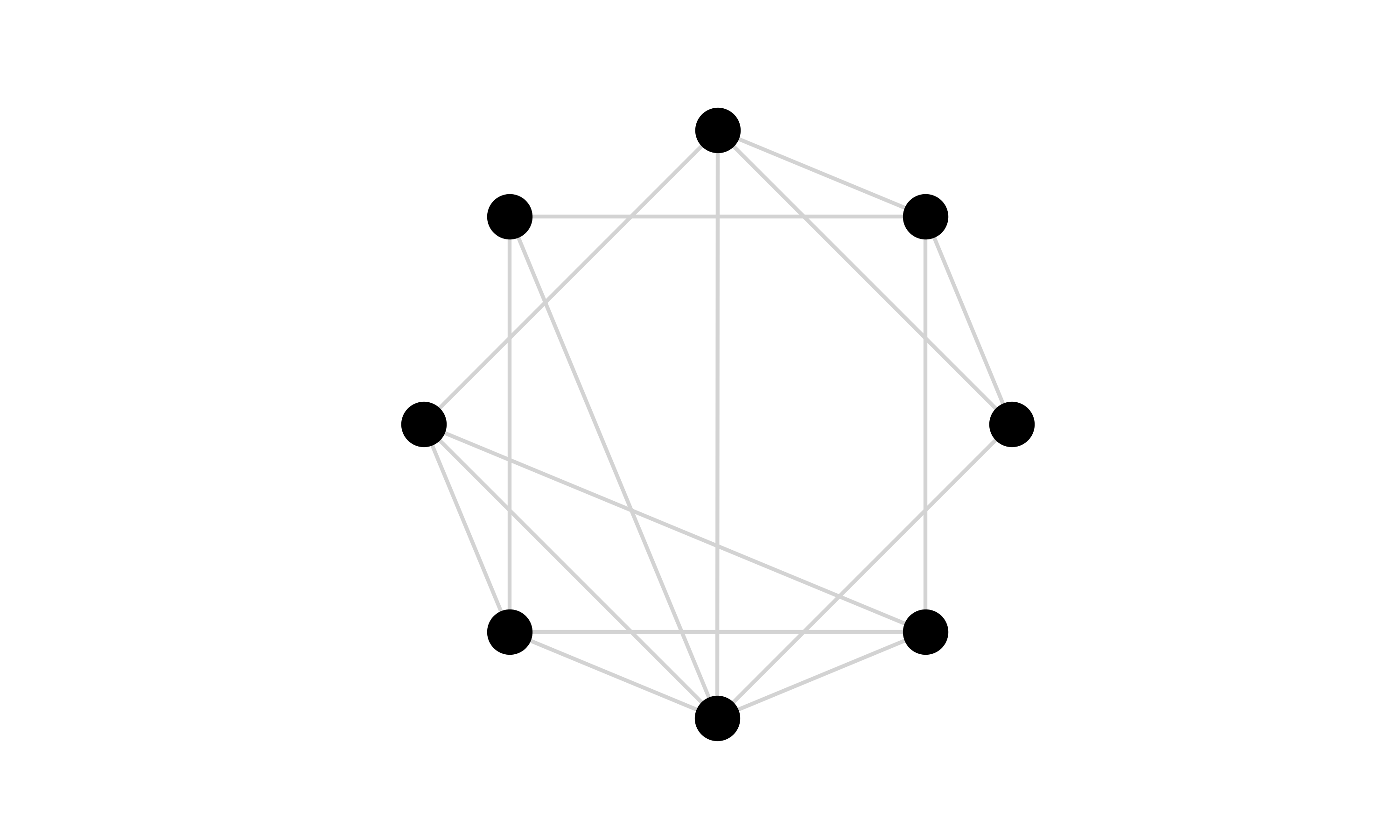}
  \caption{\centering idealized small world network}
\end{subfigure}
\begin{subfigure}{0.32\textwidth}
  \includegraphics[width=\linewidth, trim=80 50 80 50,
    clip]{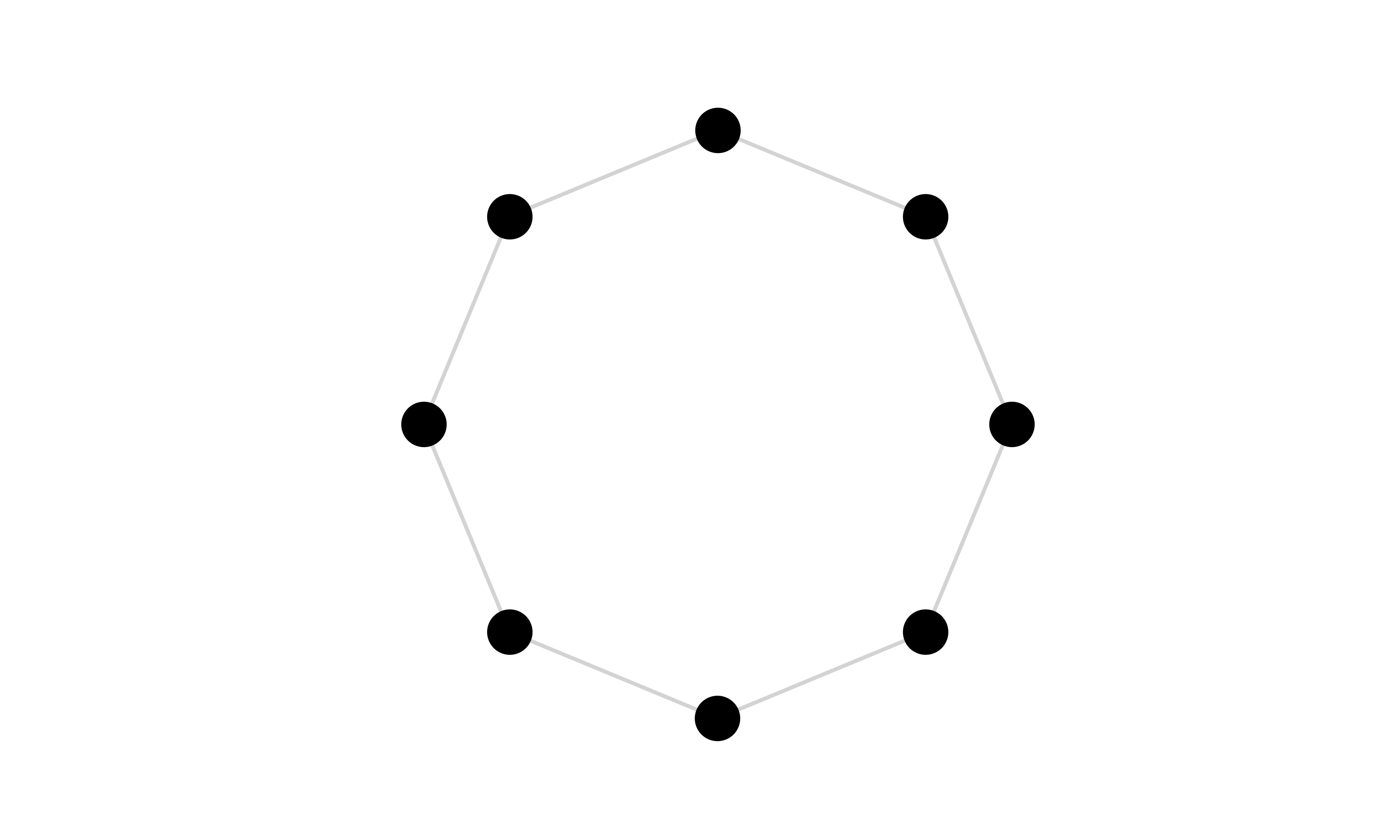}
  \caption{\centering idealized ring network}
\end{subfigure}
\begin{subfigure}{0.32\textwidth}
  \includegraphics[width=\linewidth, trim=80 50 80 50,
    clip]{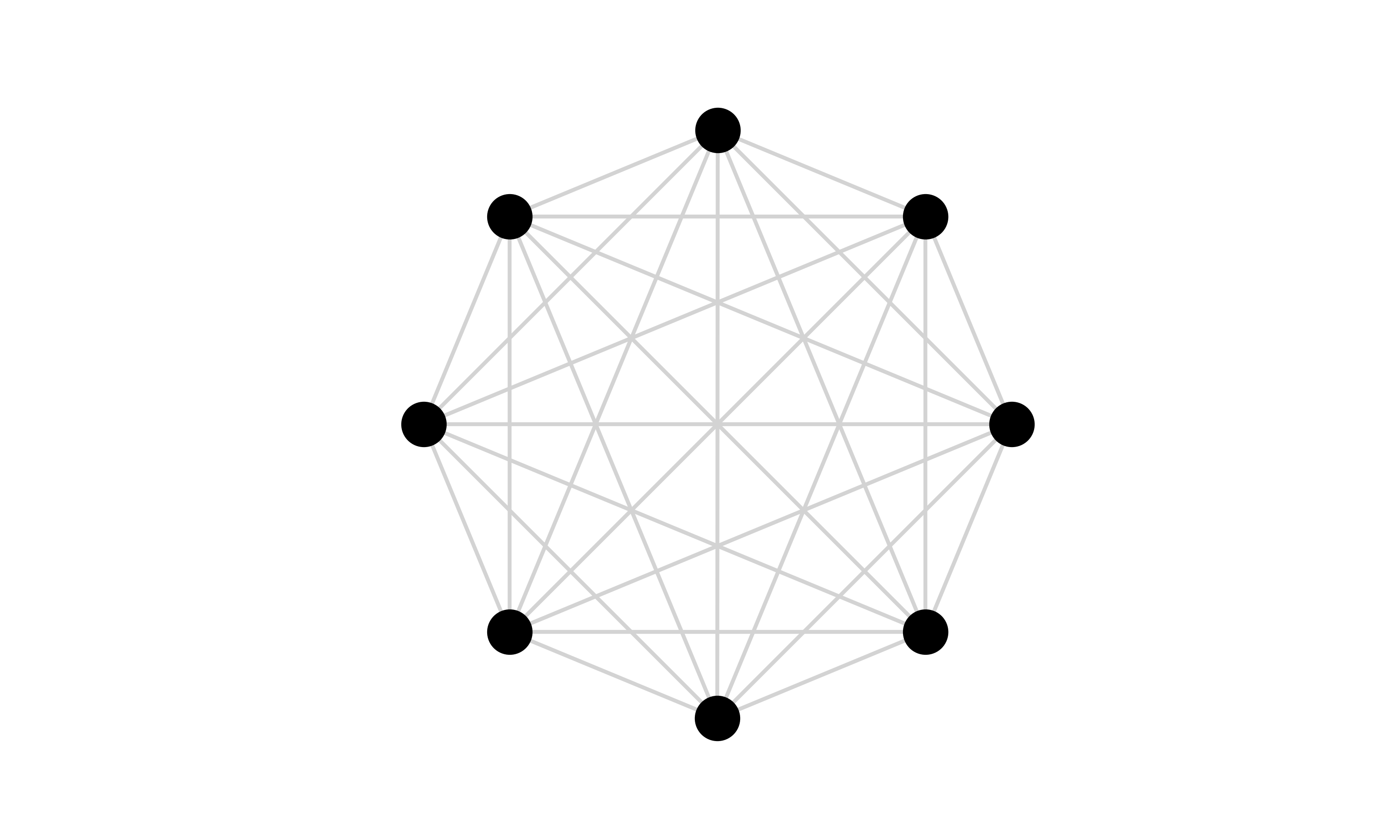}
  \caption{\centering idealized dense network}
\end{subfigure}

%\begin{subfigure}{0.32\textwidth}
%  \includegraphics[width=\linewidth]{figures/ideal_network_star}
%  \caption{\centering idealized star / wheel network}
%\end{subfigure}
%\begin{subfigure}{0.32\textwidth}
%  \includegraphics[width=\linewidth]{figures/ordinary_lowS_noAcq/%%18_MoLG_0_NA50_RS11_network}
%  \caption{\centering emergent star / wheel network}
%\end{subfigure}
%\begin{subfigure}{0.32\textwidth}
%  \includegraphics[width=\linewidth]{figures/MIA_lowS_noAcq/18_MoLG_0_NA50_RS1_network}
%  \caption{\centering MIA in star / wheel network}
%\end{subfigure}
\begin{subfigure}{0.32\textwidth}
  \includegraphics[width=\linewidth, trim=80 50 80 50,
    clip]{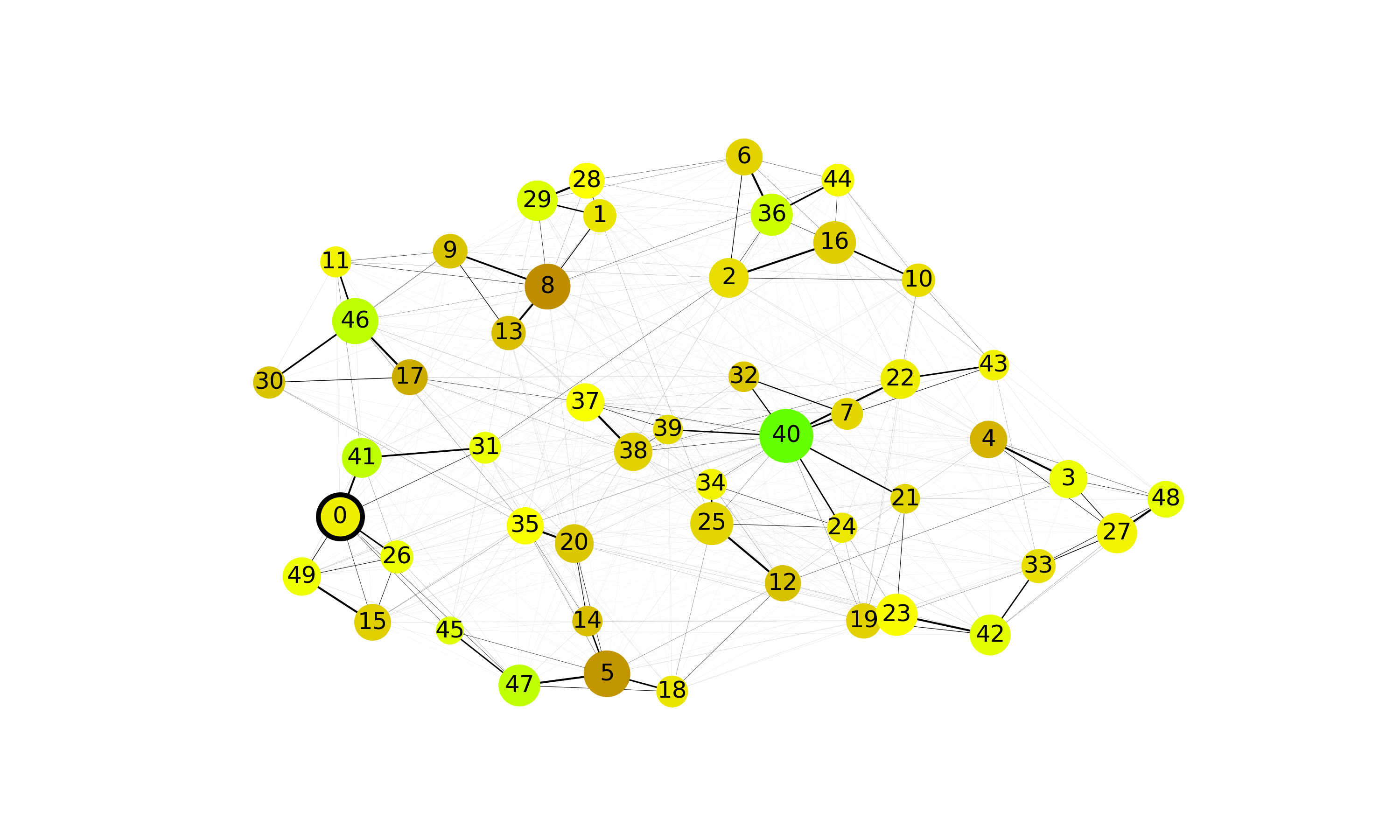}
  \caption{\centering emergent small world network}
\end{subfigure}
\begin{subfigure}{0.32\textwidth}
  \includegraphics[width=\linewidth, trim=80 50 80 50,
    clip]{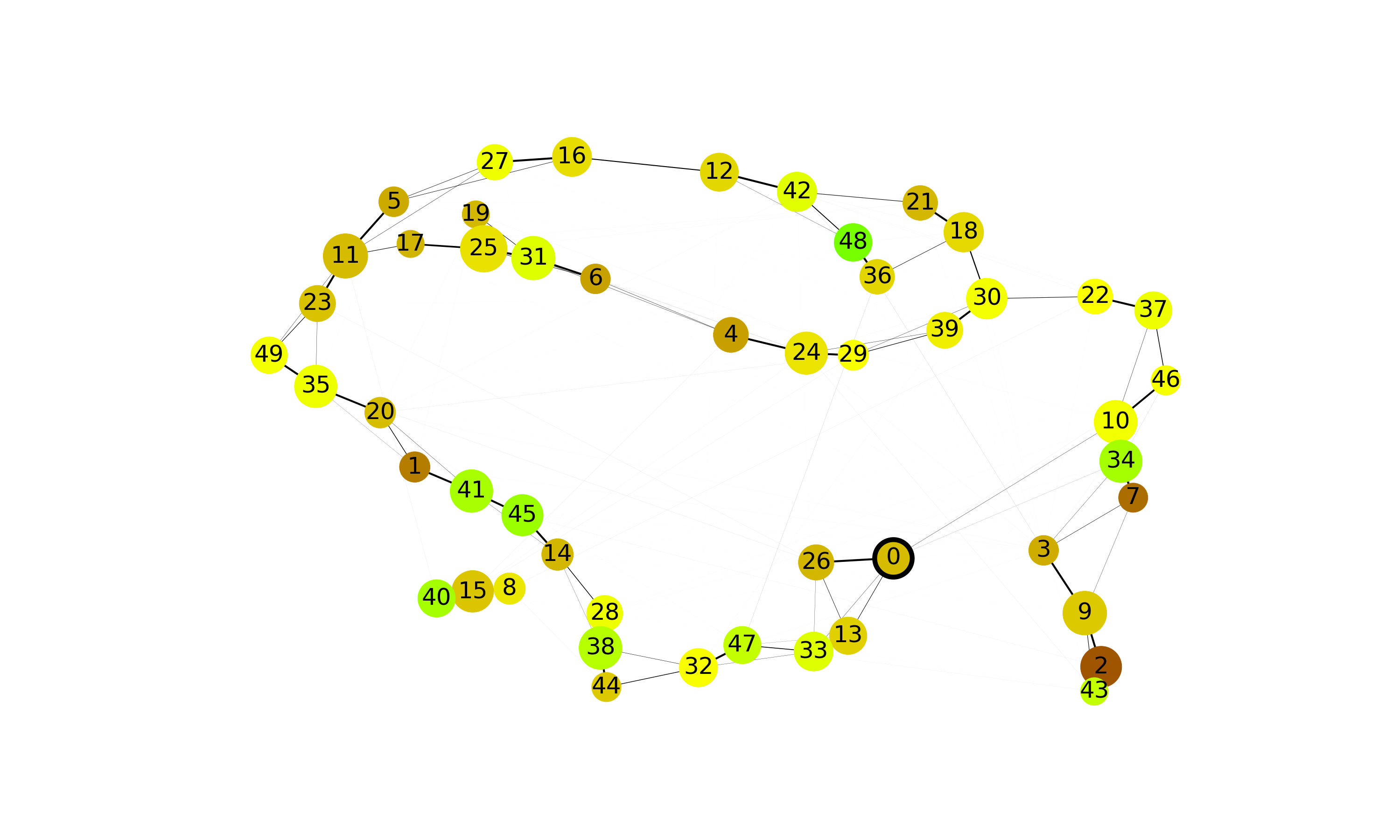}
  \caption{\centering emergent ring network}
\end{subfigure}
\begin{subfigure}{0.32\textwidth}
  \includegraphics[width=\linewidth, trim=80 50 80 50,
    clip]{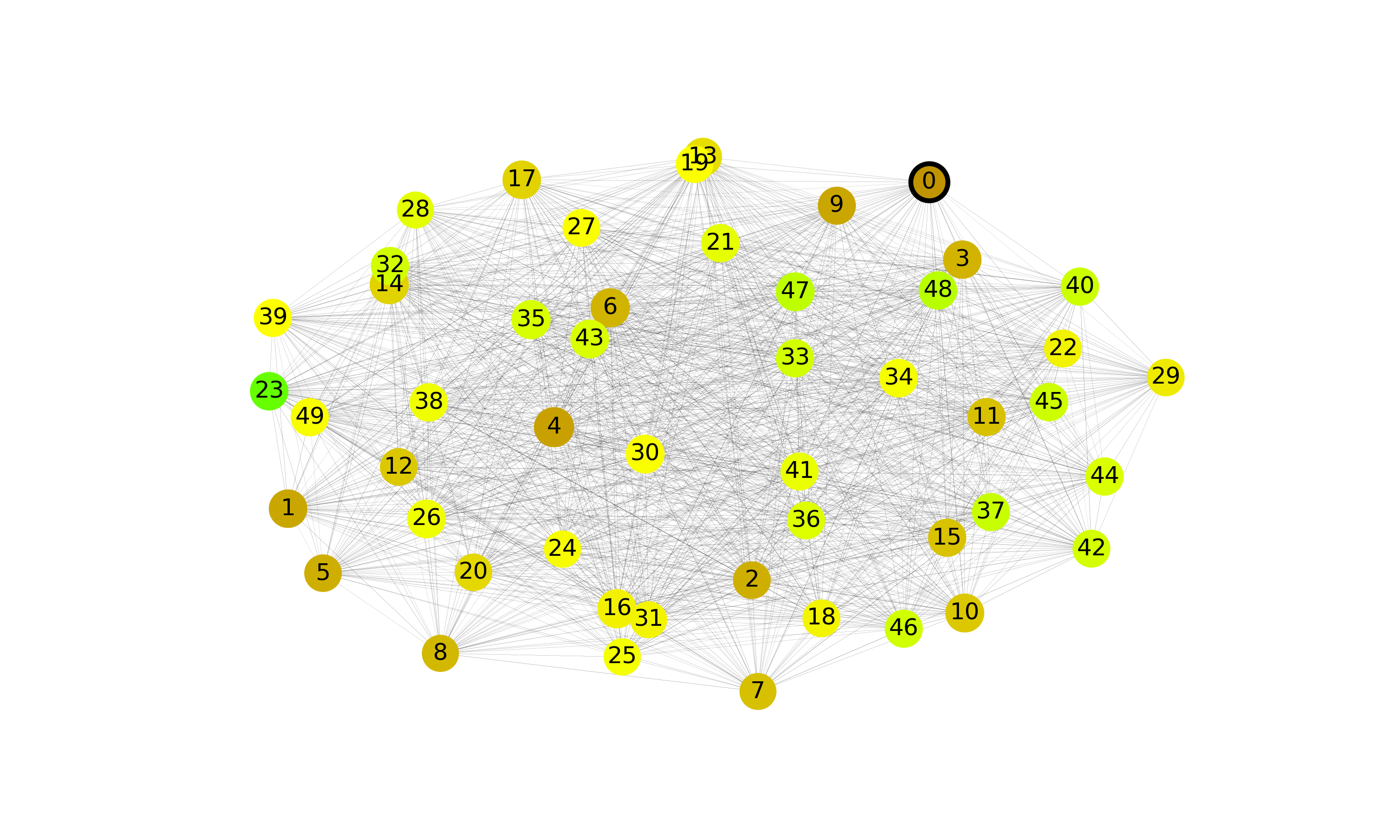}
  \caption{\centering emergent dense network}
\end{subfigure}

\begin{subfigure}{0.32\textwidth}
  \includegraphics[width=\linewidth, trim=80 50 80 50,
    clip]{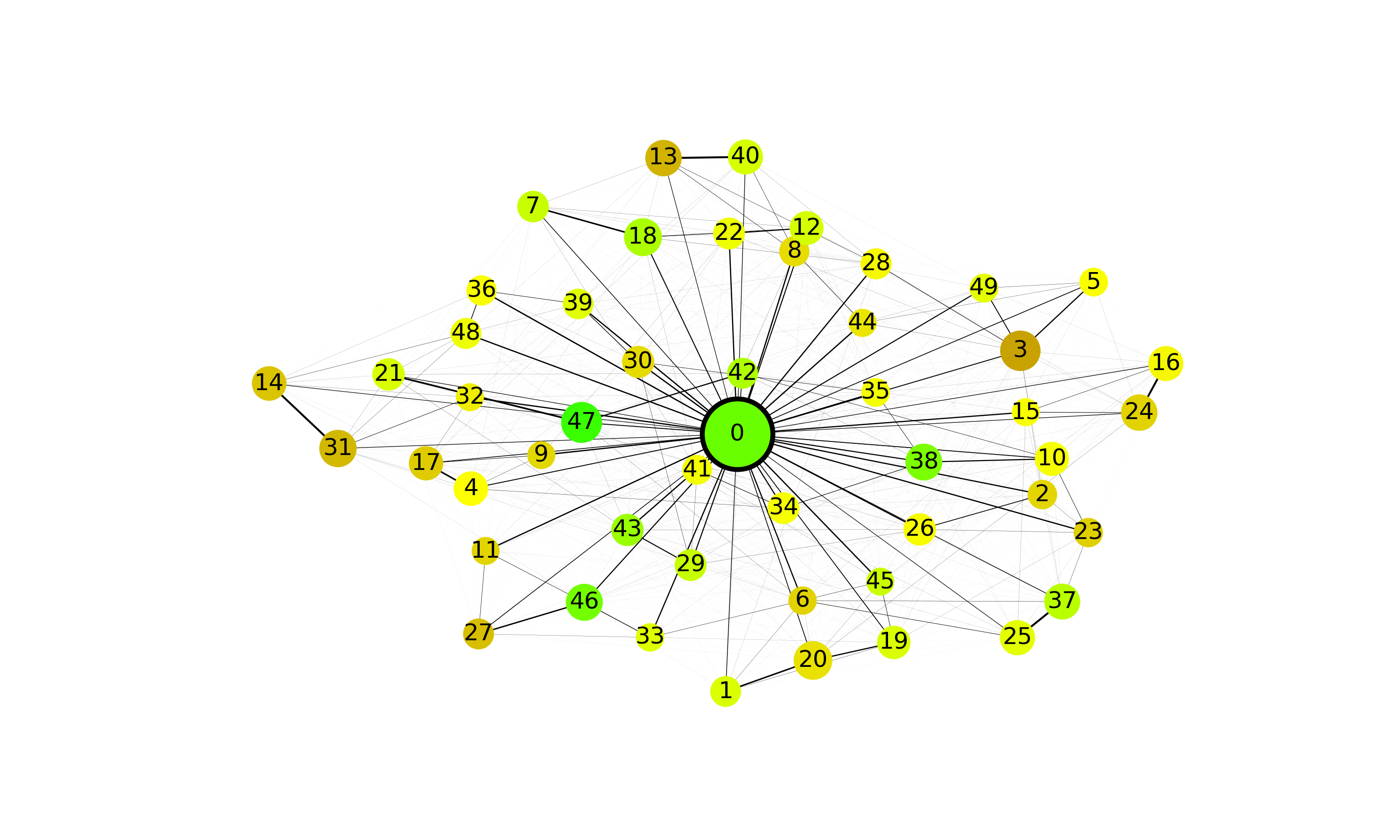}
  \caption{\centering MIA in small world network}
\end{subfigure}
\begin{subfigure}{0.32\textwidth}
  \includegraphics[width=\linewidth, trim=80 50 80 50,
    clip]{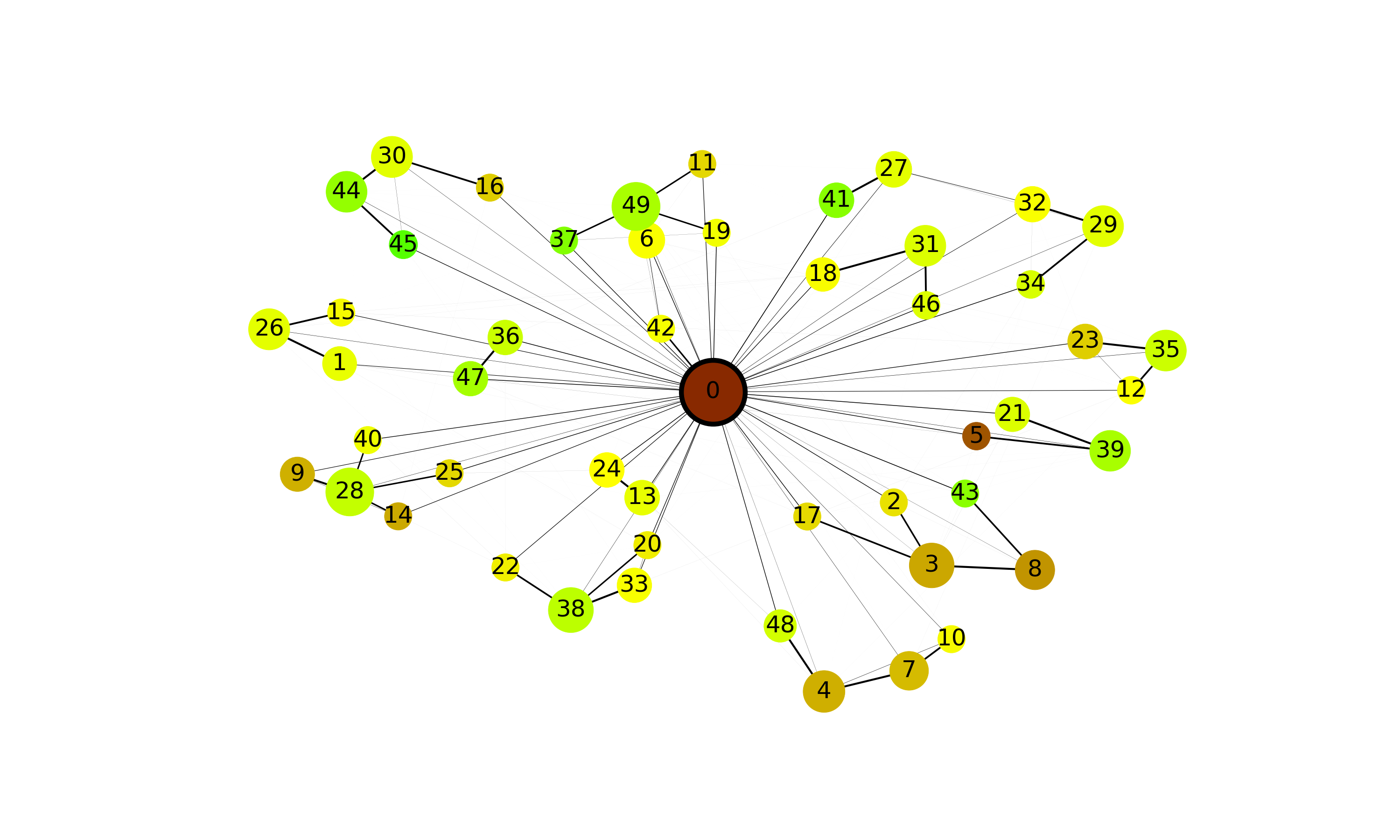}
  \caption{\centering MIA in ring network}
\end{subfigure}
\begin{subfigure}{0.32\textwidth}
  \includegraphics[width=\linewidth, trim=80 50 80 50,
    clip]{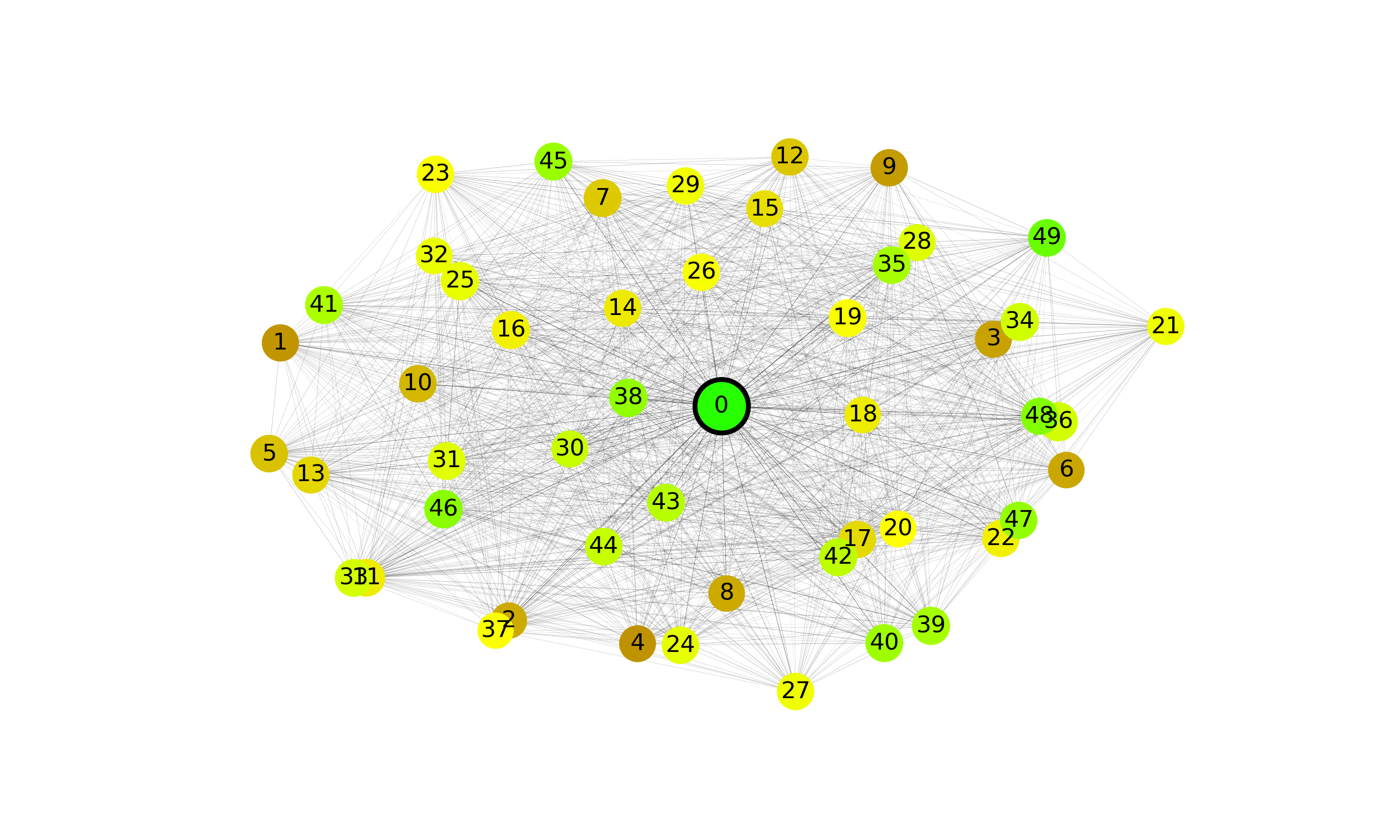}
  \caption{\centering MIA in dense network}
\end{subfigure}

\caption{Exemplary networks of single simulations with different parameter settings as given in Table \ref{tab:network_structures}. The first row shows idealized network types and the second row the realizations thereof as they emerge in the RGS with only ordinary agents. In the bottom row, we replaced ordinary agent 0 with a MIA. Colors represent agents' reputations and are shown as continuous transitions between green (high), yellow (medium) and dark brown (low). The underlying truth---agents' intrinsic honesties---can be read from their numbers (0: $0\%$ honest to 49: $100\%$ honest). Line intensities (and distances between agents as far as possible within a 2D representation) indicate how often two agents talked to each other. Node sizes indicate the degree of an agent (equation \ref{eq:degree}) in log scale.\label{fig:stability-against-mia-network-types}}
\end{figure}

Depending on the settings for ordinary agents given in Table \ref{tab:network_structures}, different network structures emerge. Examples are shown in the second row of Figure~\ref{fig:stability-against-mia-network-types}, after idealized versions in the first row. Small-world and dense networks look similar across simulations, whereas ring networks vary more. In ring setups, high shyness restricts agents to two recipients in most one-to-many interactions, producing primarily linear structures that may form circles by chance. Other linear forms, such as the loose tail in the lower right of panel \ref{fig:stability-against-mia-network-types}e, are thus also common. For studying communication dynamics, this variability is unproblematic: information typically spreads only within a small (linear) neighborhood, making it irrelevant whether its endpoints close into a circle or not.

The last row of Figure~\ref{fig:stability-against-mia-network-types} shows how the networks alter when agent 0---who is 0\% honest in any case---additionally uses a mass-influencing strategy (i.e., is changed into a MIA). All three now center around the MIA/agent 0, though there are differences between network types. In the small-world setup, almost all agents form strong ties to the MIA, reducing the relevance of local subgroups. Ring networks still form linear substructures, but their ends now all connect to the MIA, resulting in a kind of petal pattern. Importantly, not every agent has a strong tie to the MIA: many in the middle of the linear filaments have weak or no connection, a key difference from the small-world case. In dense networks, the MIA is also central, but its connections remain comparatively weak. The dense, fully connected structure mostly persists, apart from a slight overall strengthening of ties to agent 0, placing it in the center.

Moreover, the MIA achieved a very high reputation in the small-world and dense network (light green central node in \ref{fig:stability-against-mia-network-types}g,i), but only a low reputation in the ring network (dark brown central node in \ref{fig:stability-against-mia-network-types}h). While these single simulation runs are only illustrative, they still give a first impression of the MIA's performance, which will be analyzed more thoroughly in the next section.

\subsection{Propaganda with Acquaintance-Based Topic Selection (Condition A)}
\label{subsec:reputation-dynamics}
\subsubsection{Performance}

\begin{figure}[H]
\centering
\includegraphics[width=0.32\textwidth]{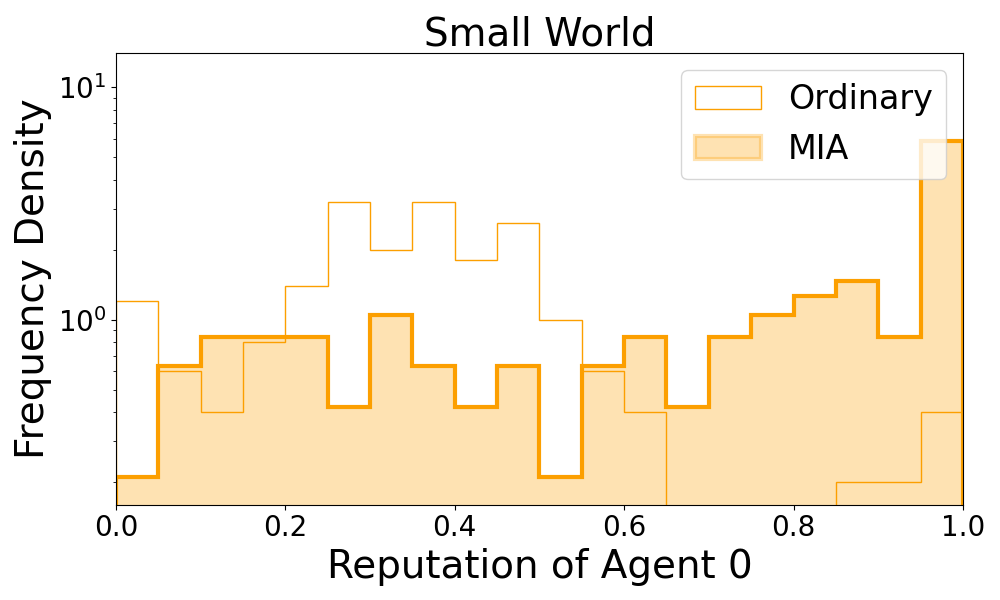}
\includegraphics[width=0.32\textwidth]{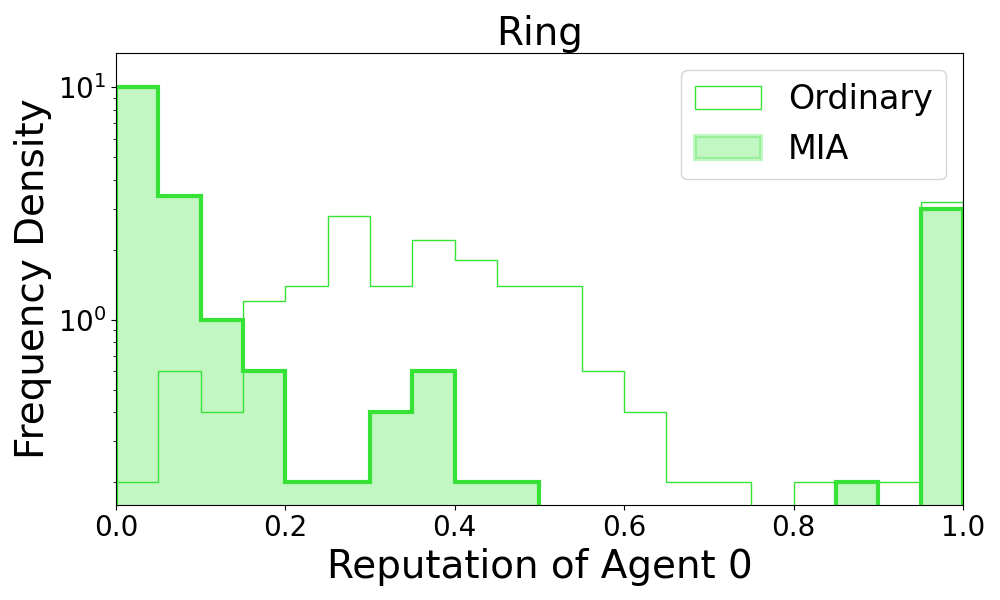}
\includegraphics[width=0.32\textwidth]{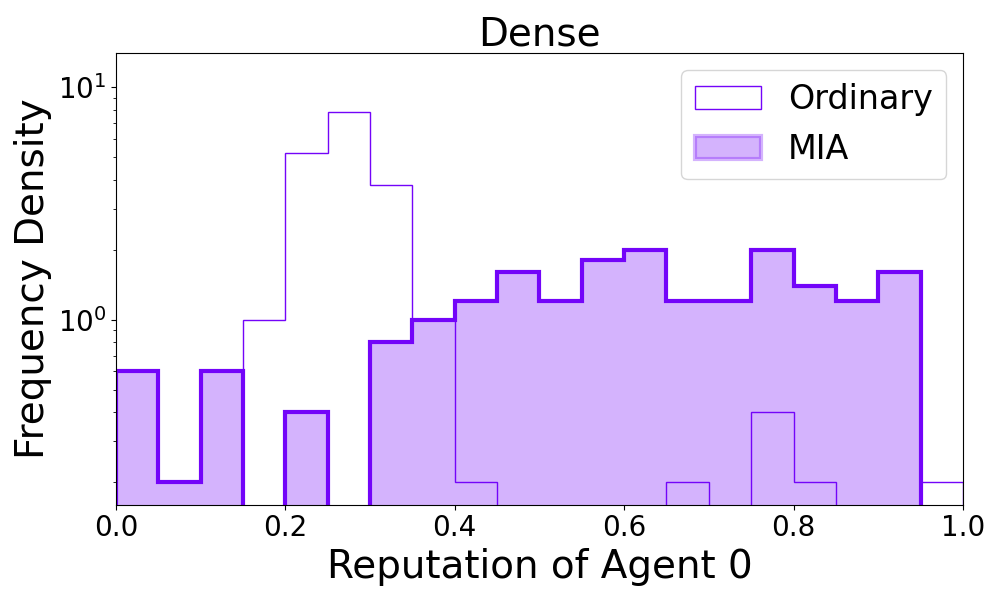}

\caption{Distribution of agent 0's reputations as seen by all others at the end of simulation runs, for each network type. Agent 0 can either be an ordinary agent (solid line) or a MIA (filled histogram).}
\label{fig:rep-hist-AcqC}
\end{figure}

We first analyze how effectively the MIA can spread its propaganda, boosting its reputation while actually being $0\%$ honest. Figure~\ref{fig:rep-hist-AcqC} shows the distribution of agent 0's ultimate reputation in the eyes of all other agents (see equation \ref{eq:reputation_group}) in all three network types, with each panel contrasting an ordinary agent 0 with a MIA version (note that an ordinary agent 0 is also always 0\% honest, like the MIA). 

With the ordinary strategy, agent 0's reputation typically centers around  around $0.4$ regardless of network type, already far above its true honesty of $0$. This effect has been observed previously in \cite{kainz2022information} and arises because agent 0, being fully dishonest, can deploy many efficient lies that go unnoticed, supporting its self-promotion\footnote{Summarizing findings from \cite{kainz2022information}: A fully dishonest agent never says its true opinion. Thus, it is nearly impossible for others to judge its honest opinion, making it harder for them to properly lie about it (as lies are tactical distortions of their honest opinions). In turn, the deceptive agent is better able to distinguish others' lies from honest statements, which makes it better informed than the average agent. With this knowledge, the deceptive agent can spread (self-promoting) lies very efficiently, such that others believe the content without noticing the dishonesty.}. In ring networks, ordinary agent~0s can reach extremely high reputations, close to $1$. However, as ordinary agents typically have only a few connections in ring structures (cf.\ Figure~\ref{fig:stability-against-mia-network-types}e), such high reputations remain confined to a small neighborhood and do not scale---hence not achieving widespread propaganda.

Due to its large outreach, the Mass Influencing Agent's reputation reflects the average perception of the entire network, not just a few neighbors. In small-world networks, the MIA  often attains reputations above $0.95$ and commonly above $0.6$. In contrast, the MIA in a ring network performs less reliably: while it sometimes achieves top reputations, it also has a high risk of falling below $0.2$ (worse than than an ordinary agent 0). This hints at the existence of a critical factor that determines whether propaganda succeeds spectacularly or backfires. \S\ref{subsubsec:centrality-and-reputation} will trace this back to the MIA's centrality. 

In dense networks, the MIA performs reliably well, typically achieving medium to high reputations ($0.5$ to $0.95$) with only a negligible chance of doing worse than ordinary agents. However, top reputations above $0.95$ are rare, unlike in small-world and ring networks. This difference lies in a self-reinforcing effect that appears in small-world and ring structures after a certain threshold but cannot in dense networks (a mechanism analyzed in \S\ref{subsubsec:centrality-and-reputation}).

In general, propaganda can be very successful in all three networks, but for fundamentally different reasons and in different ways. This distinction becomes clear from the networks' structural responses to the MIA's behavior and the resulting opinion dynamics, discussed below.

\subsubsection{Causes \& Underlying Mechanics}

\paragraph{High Centrality}\label{subsubsec:centrality-and-reputation}

\begin{figure}[H]
\centering
\includegraphics[width=0.32\textwidth]{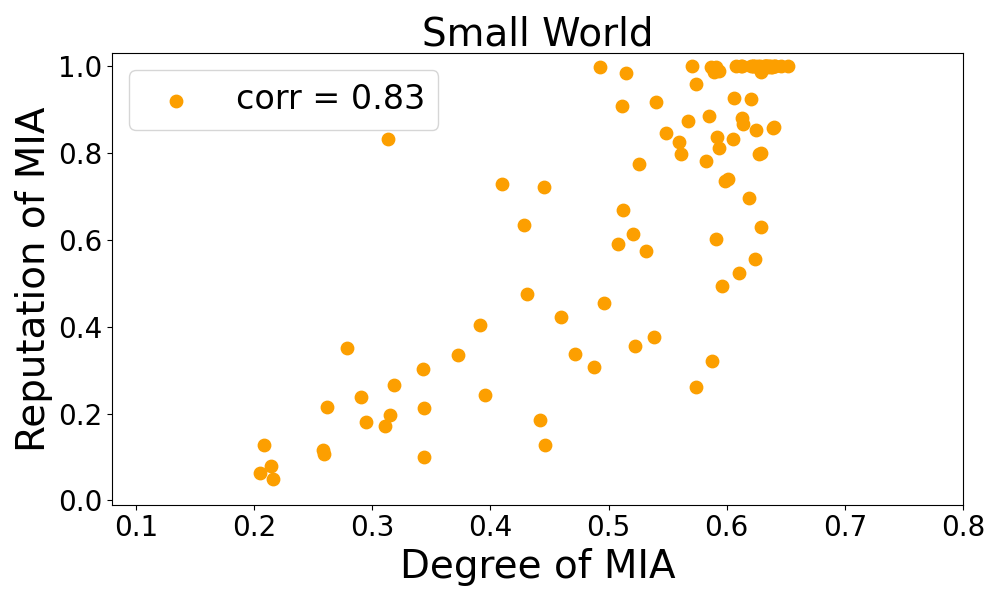}
\includegraphics[width=0.32\textwidth]{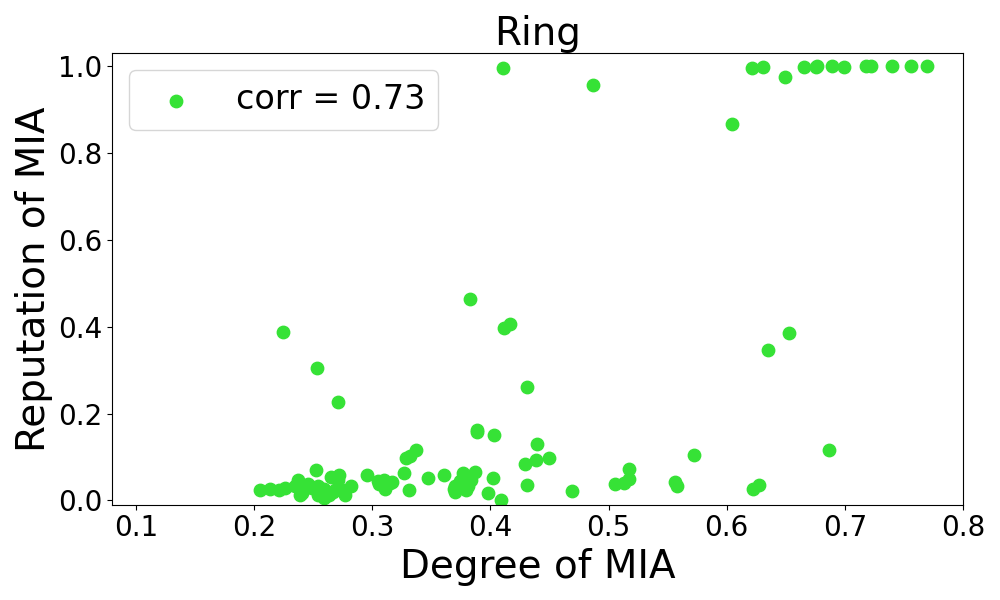}
\includegraphics[width=0.32\textwidth]{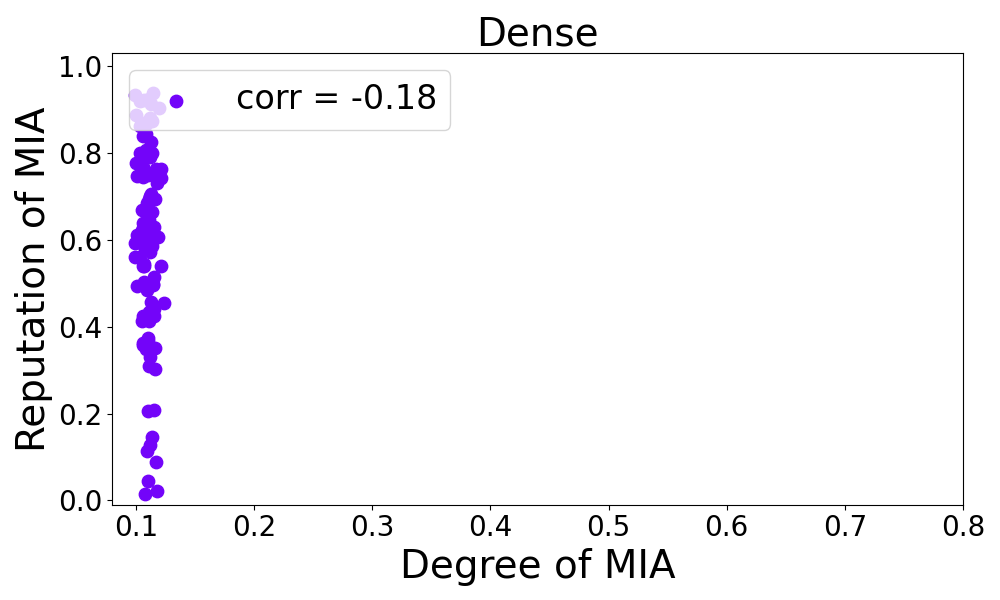}

\caption{Each point shows a MIA's reputation and its degree at the end of a single simulation run, with panel insets showing the corresponding correlation across all runs within a given network type.}
\label{fig:degree-rep-AcqC}
\end{figure}

The emerging network structure is the main driver of communication and opinion dynamics, with MIA's centrality largely determining whether its propaganda succeeds or fails. Figure~\ref{fig:degree-rep-AcqC} shows the correlation between MIA's degree and reputation, where degree is a weighted, normalized measure of how many conversations an agent participates in (basically, a weighted version of the commonly known degree centrality, cf.\ equation~\ref{eq:degree}). 

In both small-world and ring networks, a higher degree is strongly correlated with a higher reputation. In ring networks, reputation even collapses sharply once the degree drops below $0.6$, reflecting the same success–failure threshold seen in Figure~\ref{fig:rep-hist-AcqC}: a lower degree means fewer interactions involving the MIA and more interactions between ordinary agents (reflected in longer linear filaments in the petal pattern illustrated in Figure~\ref{fig:stability-against-mia-network-types}h, where the MIA has little influence on longer petals' extremes). High degree means shorter such filaments, giving the MIA direct influence on more individuals. Thus, in rings, the MIA succeeds only when these filaments stay short; otherwise its reputation falls immediately. In small-world networks the relation is smoother because even weakly connected agents still retain some access to the MIA.

\begin{figure}[H]
\centering
\includegraphics[width=0.32\textwidth]{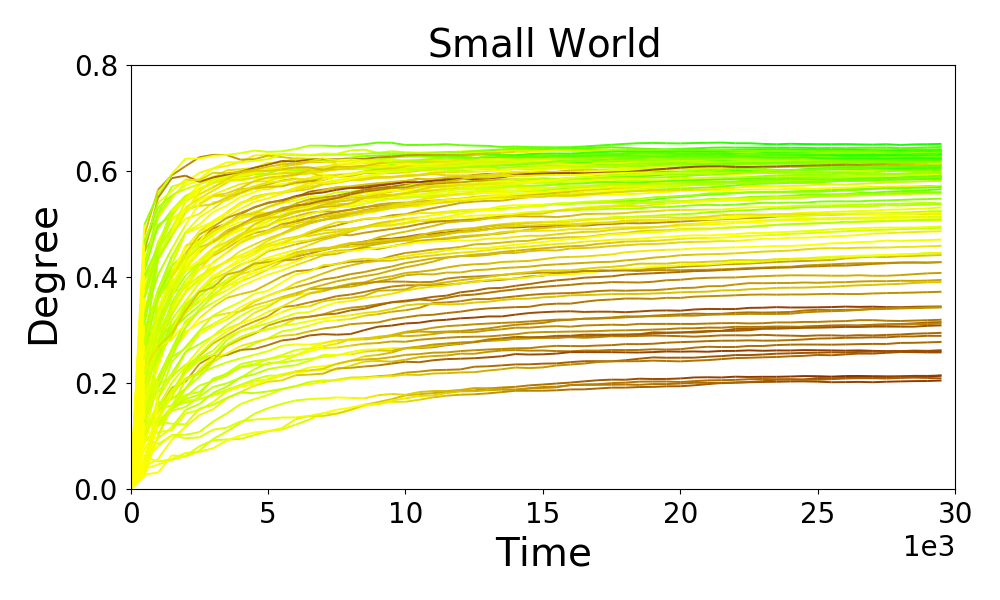}
\includegraphics[width=0.32\textwidth]{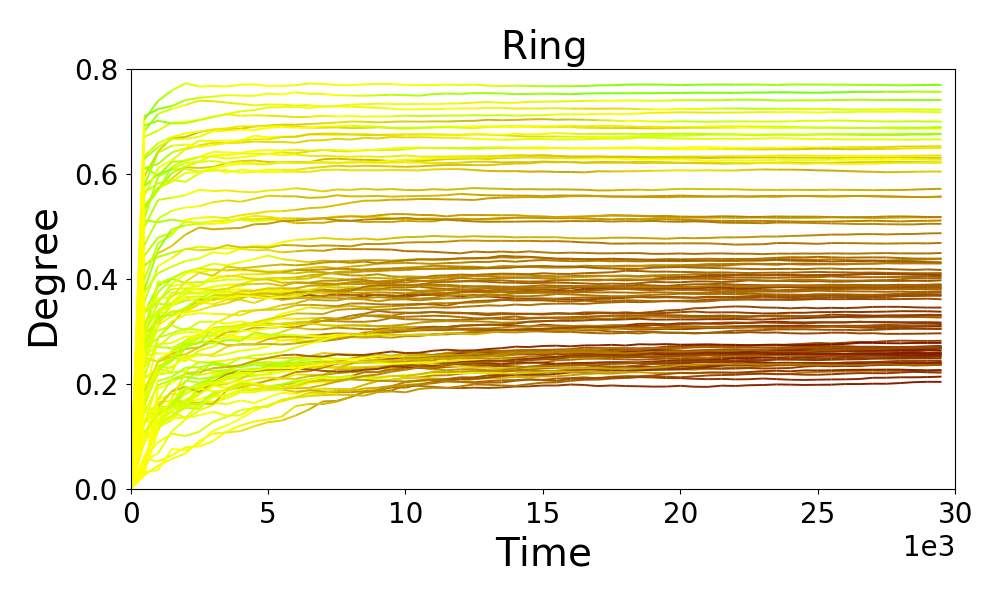}
\includegraphics[width=0.32\textwidth]{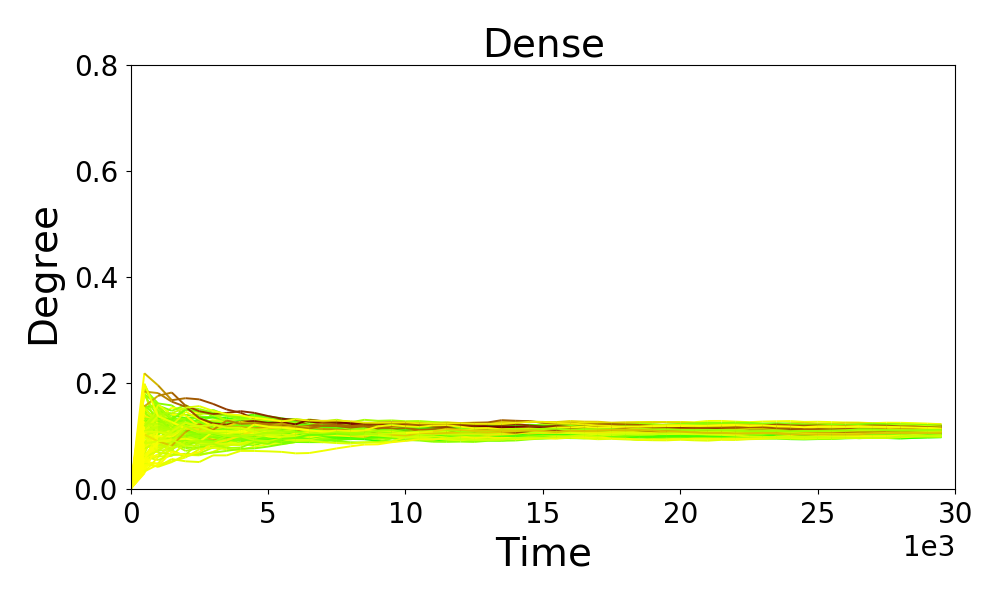}

\caption{Time evolution of the MIA's degree, for each of the three network types, with lines connecting data from a single simulation run. Line color indicates a MIA's reputation at any point in time, ranging from dark brown ($0\%$) over yellow ($50\%$) to green ($100\%$).}
\label{fig:degree-time-AcqC}
\end{figure}

Dense networks behave fundamentally differently: the MIA's degree remains consistently low, so degree and reputation are essentially uncorrelated. Because opinions depend on interactions with constantly changing partners---mostly ordinary agents talking to other ordinary agents---opinion formation becomes highly randomized and cannot be attributed to the MIA's direct connections alone. This is also visible in the time evolution of the MIA's degree (right panel of Figure~\ref{fig:degree-time-AcqC}), which quickly stabilizes around $0.11$ in dense networks, regardless of reputation. 

In small-world and ring networks, the MIA's final reputation depends both on its final degree and---crucially---on its early structural position. Reputation begins at a moderate level in most runs but then diverges: in ring networks the final degree is reached quickly, and reputation (directly correlated with it) settles shortly afterward. Small-world networks show a more intricate pattern. High degrees quickly lead to lower reputations because the MIA cannot exploit its full power on the whole network at once. Over time, however, broad outreach pays off, and high-degree MIAs eventually surpass those with small degrees in reputation. The latter maintain their early high reputation only briefly; the rest of the network ultimately counterbalances and corrects the initially targeted subgroup.

Overall, the early network structure the MIA forms determines its final reputation: the larger the targeted crowd, the higher the eventual reputation in both small-world and ring networks.

\paragraph{Instability \& Polarization}\label{subsubsec:group-vs-individuals-opinions}

When we take a closer look at how volatile ordinary agents' opinions about the MIA are, rather than just considering the MIA's average reputation, several new patterns emerge. Volatility here reflects the variability in one agent's estimate of another's reputation within a rolling time window, which helps explain why propaganda can take hold in some cases but fails in others.

\begin{figure}[H]
\centering
\includegraphics[width=0.32\textwidth]{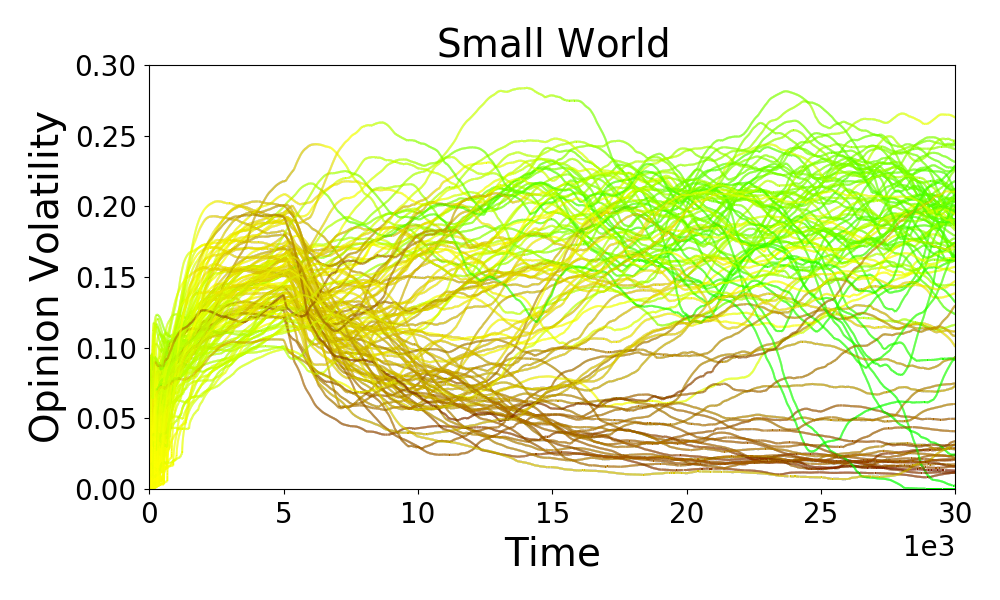}
\includegraphics[width=0.32\textwidth]{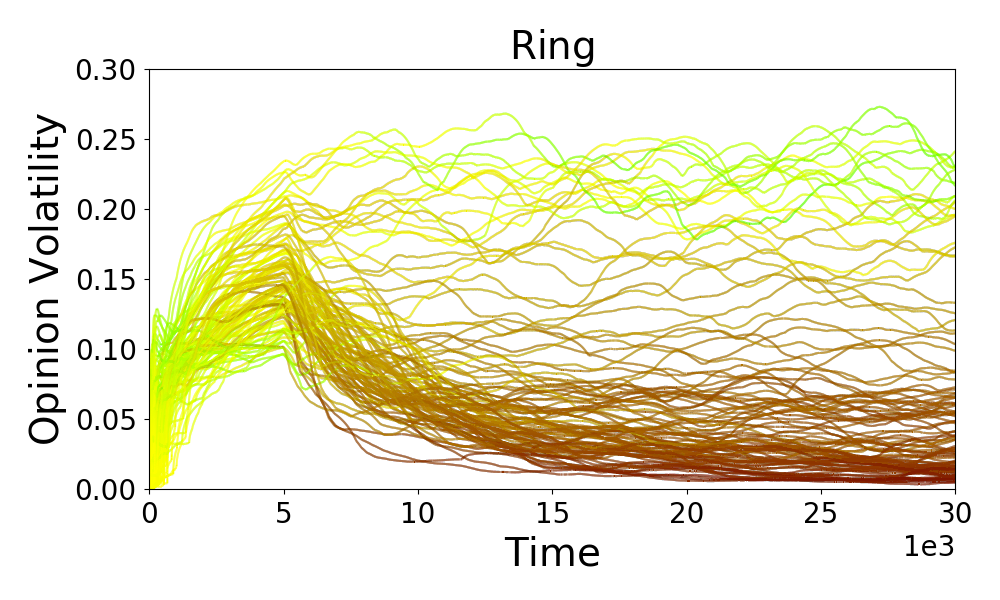}
\includegraphics[width=0.32\textwidth]{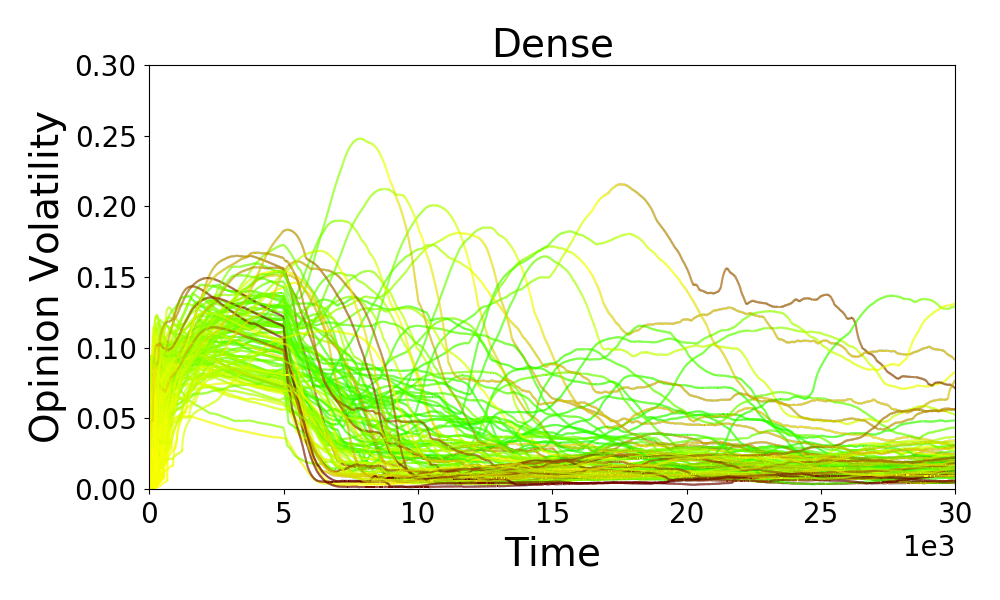}

\caption{Evolution over time of the average volatility of ordinary agents' opinions, for each of the three network types. Individual volatilities are calculated following equation \ref{eq:volatility} using a rolling window with a size of $T=5000$ timesteps. The color indicates the MIA's reputation at any point in time, ranging from dark brown ($0\%$) over yellow ($50\%$) to green ($100\%$).}
\label{fig:opinion-volatility-AcqC}
\end{figure}

Figure~\ref{fig:opinion-volatility-AcqC} shows all agents' average volatility over time, including how this relates to average reputation of the MIA. When lines are low on the y-axis, individual opinions barely change; where lines are high on the y-axis, individual opinions are very unstable. For both small-world and ring networks, when agents follow the propaganda, the MIA enjoys a high average reputation even while the individual opinions feeding into this average are volatile. Further, especially for the small-world network, these even cluster somewhat, appearing as two loose ``branches'' of the evolutionary trajectory: one in which agents barely change their minds (lower branch: stable, poor MIA reputation), and another in which they shift repeatedly (upper branch: volatile, high MIA reputation). 

\begin{figure}[H]
\centering
\includegraphics[width=0.32\textwidth]{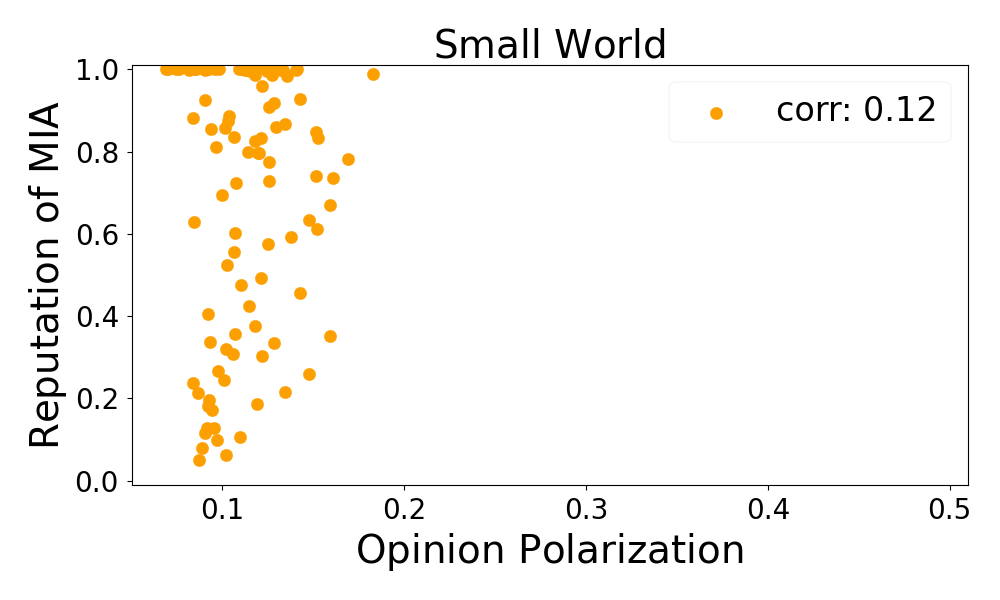}
\includegraphics[width=0.32\textwidth]{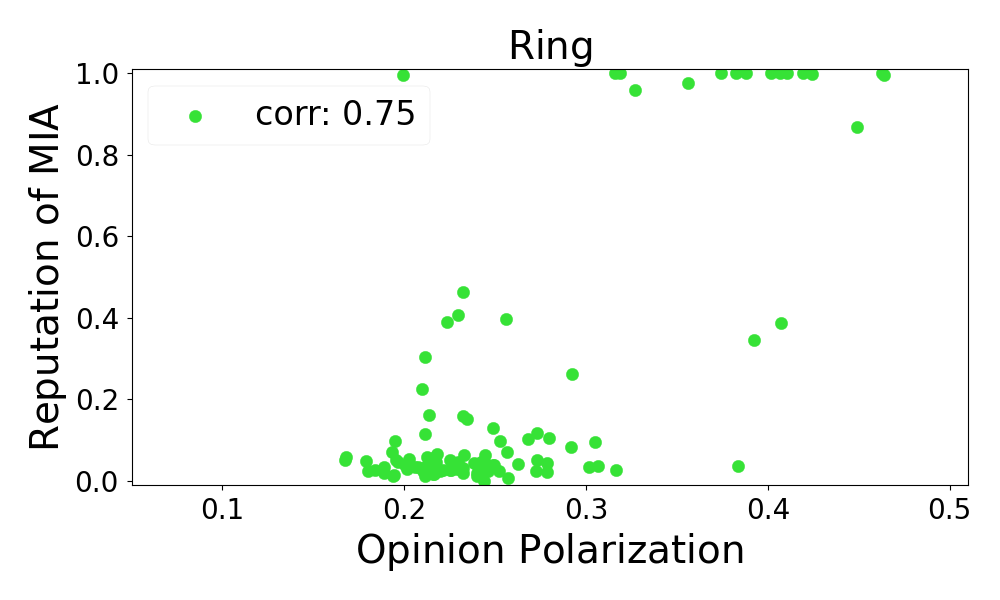}
\includegraphics[width=0.32\textwidth]{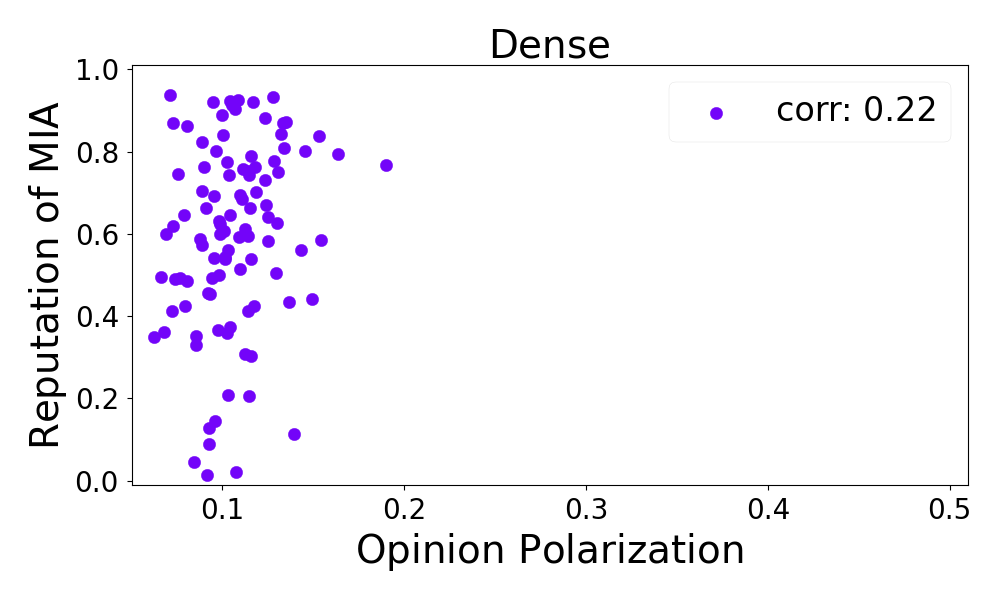}

\caption{Each point shows a MIA's reputation and how polarized the network is about that reputation at the end of a single simulation run, with panel insets showing the corresponding correlation across all runs within a given network type.}
\label{fig:polarization-AcqC}
\end{figure}

Looking at the polarization of opinions about the MIA\footnote{Although the polarization measure (equation \ref{eq:polarization}) is defined to capture overall group polarization, in this setup, group polarization mainly reflects polarization about the MIA. This is because all agents frequently encounter the MIA and have strong opinions about it, whereas ordinary agents are only known locally with few other agents having an informative opinion about them.}, we further observe a split of this high-reputation, high-volatility regime in small world and ring networks into two distinct patterns (Figure~\ref{fig:polarization-AcqC}). In small-world networks, polarization stays low: even volatile agents generally align with the group, and opinion shifts arise mainly as short-term reactions from direct observations of the MIA giving off a tell, and thus behaving surprisingly from the perspective of agents who view it as honest. 

In ring networks, however, high reputation comes with high polarization. Here only the agents close to the MIA share a positive opinion of it, while those shielded by long filaments remain unconvinced. This structural separation explains why the MIA performs less reliably in rings: evidence hinting towards a low reputation of the MIA comes not only from personal observations but also from exchanges with other ordinary agents who are not in direct contact with the MIA, causing high polarization between the ones closely connected to it and the ones who are not.

In dense networks we observe neither such volatility split nor polarization differences (right panels of figures~\ref{fig:opinion-volatility-AcqC} and~\ref{fig:polarization-AcqC}). Opinions are consistently stable, and even high reputations of the MIA emerge slowly, steadily, and with network-wide consensus. Whatever dense networks agree on, they reach collectively.

\paragraph{Echo Chambers}
Bringing these threads together, we can now understand what allows the MIA to reach extremely high reputations in small-world and ring networks, but not in dense ones. In small-world and ring networks, self-reinforcing feedback loops form when agents repeatedly confirm each others' views in small subgroups while being highly engaged with the MIA at the same time. Because partner choice is steered by acquaintance, the MIA is selected disproportionately often by ordinary agents and thus gains many more opportunities to interact (and inject propaganda) than any other agent. Small and tightly focused echo chambers are the consequence. Once a critical threshold of influence is crossed, this mutual reinforcement escalates rapidly.

Dense networks cannot sustain such dynamics: their constant mixing of diverse partners prevents prolonged mutual confirmation and limits the MIA's presence. Propaganda can still spread, but only gradually and without the runaway amplification seen in the other two network types, causing the strong peaks at high reputations earlier in figure \ref{fig:rep-hist-AcqC}.

\subsubsection{Interim Conclusion --- Condition A}

Although small-world, ring, and dense networks differ strongly in how propaganda spreads, they all end up giving almost their full attention to the MIA. Whether through structural centrality (small-world), dominance along linear filaments (ring), or sheer interconnectedness (dense), the MIA becomes the focal point of communication, making strong propaganda effects possible in every case.

What happens if this attention is no longer concentrated on a single topic? The next section will show how randomized topic choice can actually foster resilience.

\subsection{Propaganda with Randomized Topic Selection (Condition B)}
\label{subsec:effects-of-randomized-topic-choice}

To prevent the entire network from focusing solely on the MIA, we let agents choose conversation topics randomly instead of based on acquaintance. We revisit the same metrics from \S\ref{subsec:reputation-dynamics} (reputation, centrality, opinion volatility and polarization), highlighting key differences from acquaintance-driven topic choice above.

\subsubsection{Resilience at the Group Level}

\begin{figure}[H]
\centering
\includegraphics[width=0.32\textwidth]{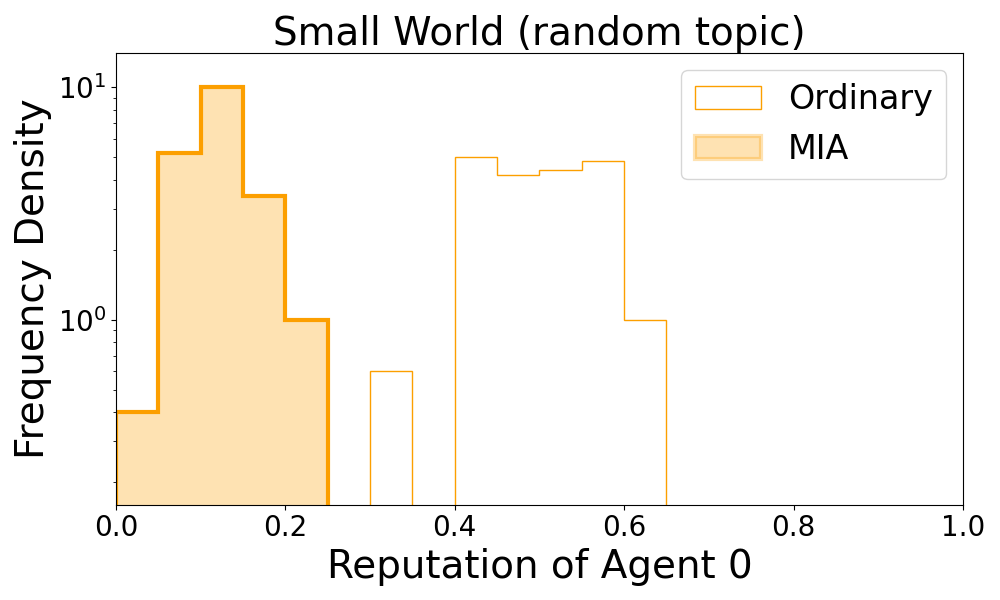}
\includegraphics[width=0.32\textwidth]{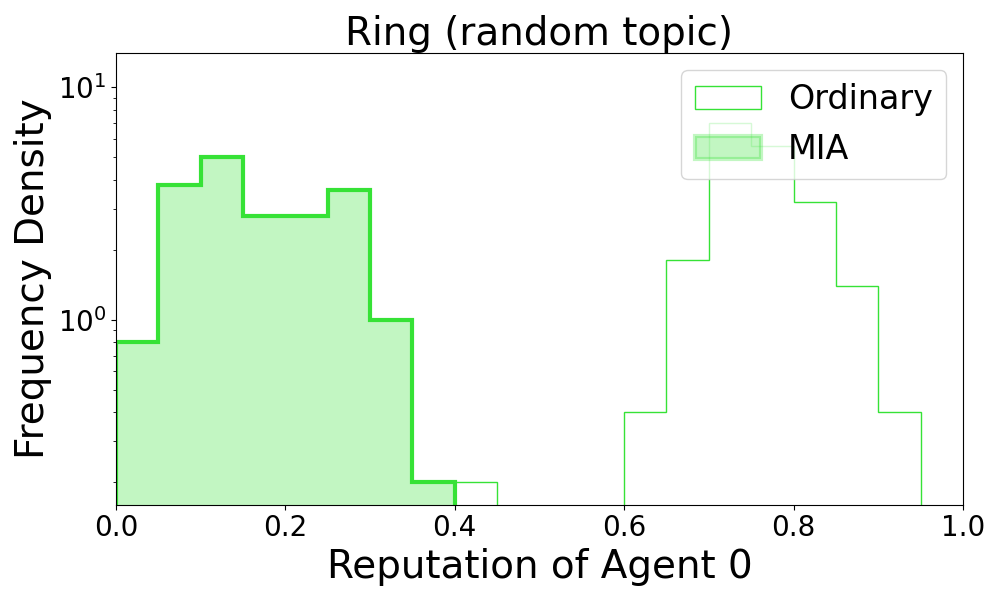}
\includegraphics[width=0.32\textwidth]{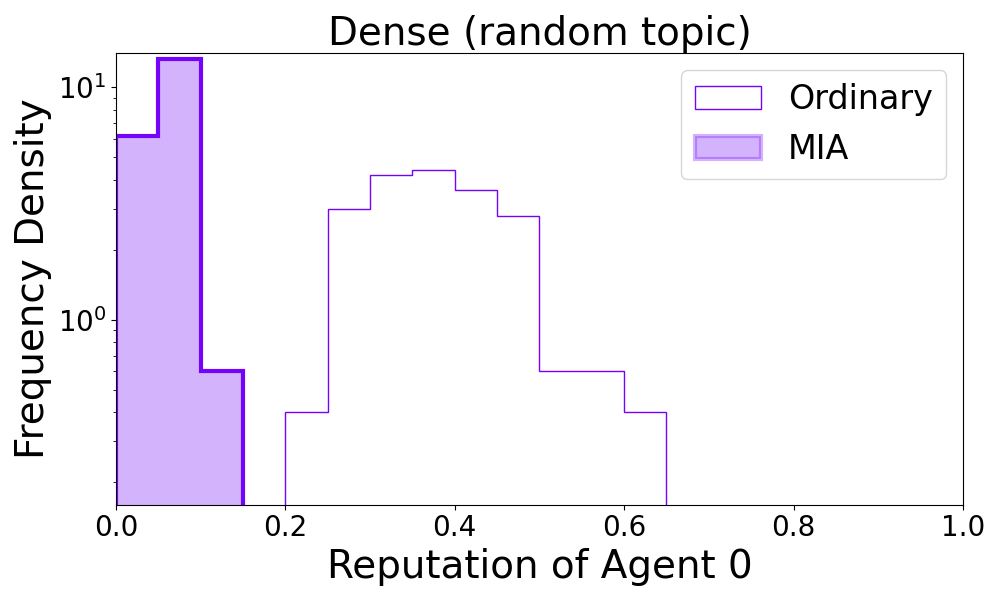}

\caption{Same as Figure \ref{fig:rep-hist-AcqC} but with randomized topic choice for all ordinary agents.}
\label{fig:rep-hist-noAcqC}
\end{figure}

With randomized topic choice, Figure~\ref{fig:rep-hist-noAcqC} shows that the MIA's reputation drops significantly across all network types, most drastically to around $0.1$ in dense and small-world networks. In ring networks, the previously sharp peak at very low reputations spreads out to around $0.2$, but the peak at top reputations also disappears completely. With random topics, the MIA's propaganda fails entirely, and this is not just about MIA's 0\% honesty, as the ordinary agent 0 is not impacted in the same way. 

Further, though forced to choose their conversation topics randomly, ordinary agents' opinions of MIA are still built on strong evidence. If not, their opinions---and thus the MIA's reputation---would stay close to the prior at $0.5$. Instead, Figure~\ref{fig:rep-hist-noAcqC} shows how, even in the dense network where both partner and topic choice are maximally random, MIA reputations cluster sharply around~$0.1$. Such a pronounced, coherent shift is only possible if agents accumulate substantial (and directionally consistent) information. Randomizing topics thus removes the MIA's propagandistic leverage while still providing agents with enough information to build sophisticated opinions, becoming resilient.

As an aside: the apparently higher reputations of ordinary agent~0 in Figure~\ref{fig:rep-hist-noAcqC} compared to Figure~\ref{fig:rep-hist-AcqC} are an artifact of poor information. Most agents simply never encounter agent 0, so opinions usually remain uninformed and near the prior. Only in ring networks do ordinary agents reach seemingly high values around $0.8$, but this stems from their tiny local neighborhood, where frequent uninformed statements and ``white lies'' make them appear honest. In networks with ordinary agents only and randomized topic choice, this state is rarely surpassed.

\subsubsection{Causes \& Underlying Mechanics}

\begin{figure}[H]
\centering
\includegraphics[width=0.32\textwidth]{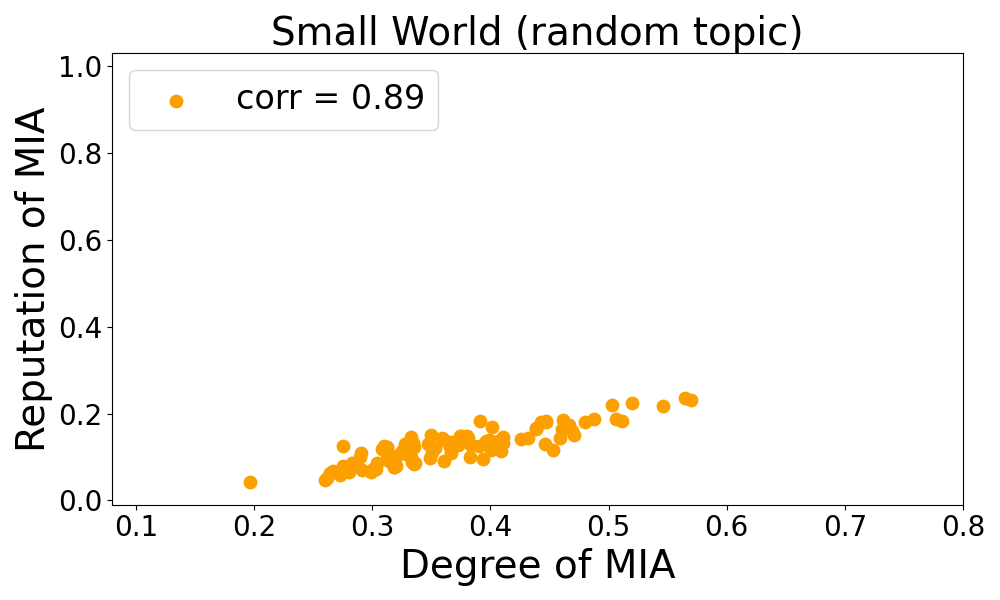}
\includegraphics[width=0.32\textwidth]{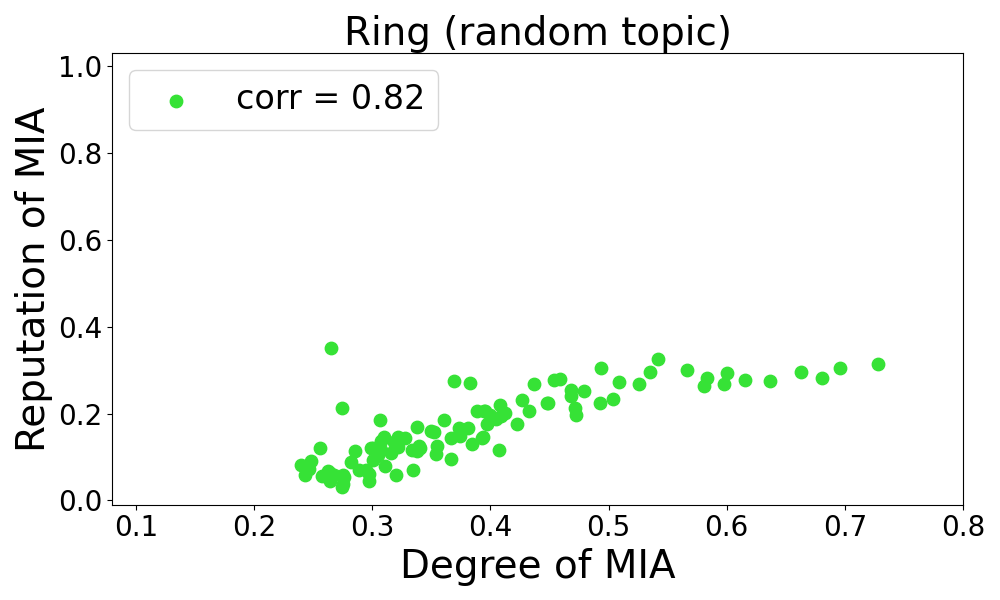}
\includegraphics[width=0.32\textwidth]{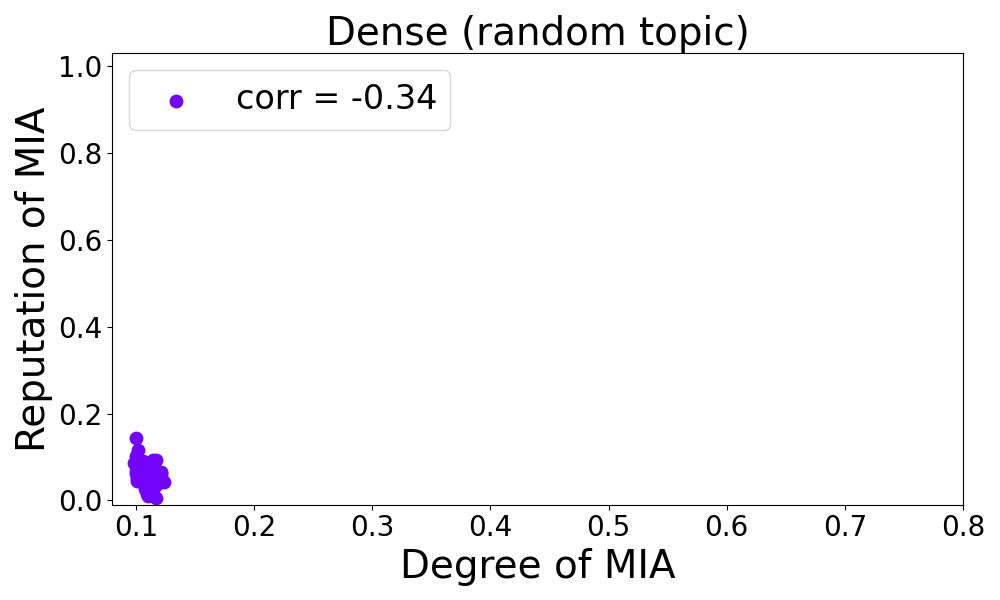}

\caption{Same as Figure \ref{fig:degree-rep-AcqC} but with randomized topic choice for all ordinary agents.}
\label{fig:degree-rep-noAcqC}
\end{figure}

\begin{figure}[H]
\centering
\includegraphics[width=0.32\textwidth]{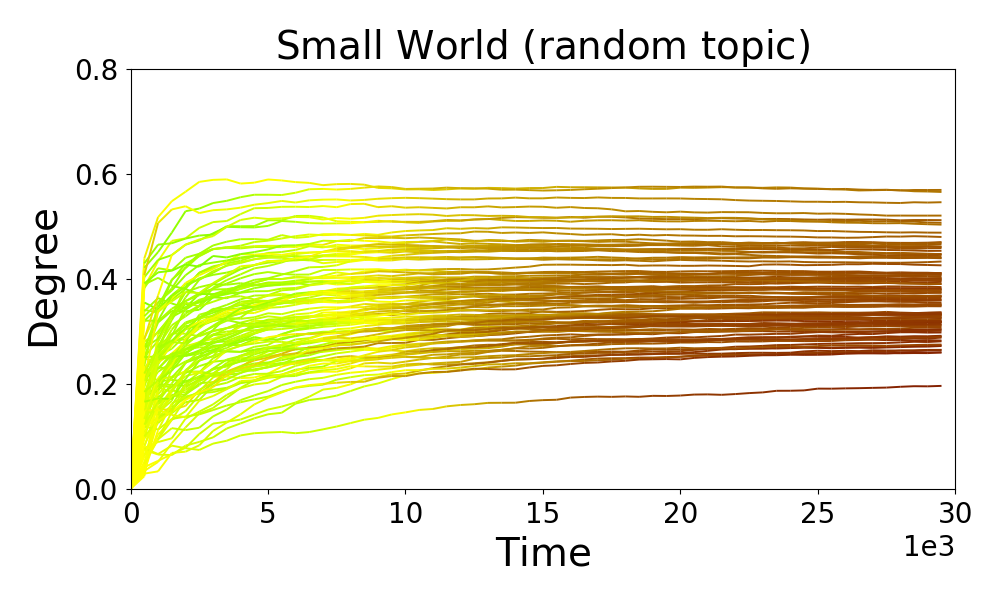}
\includegraphics[width=0.32\textwidth]{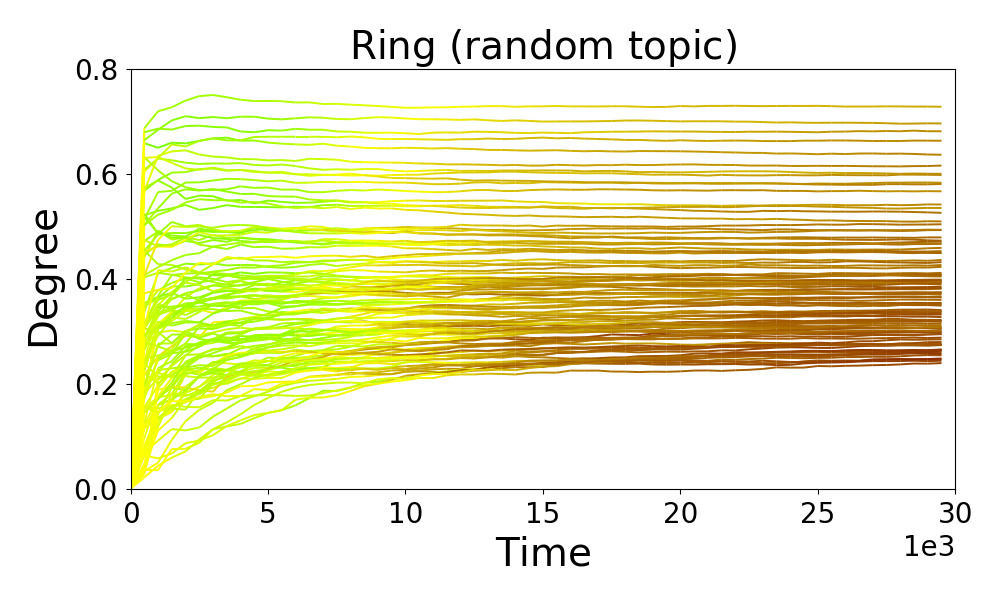}
\includegraphics[width=0.32\textwidth]{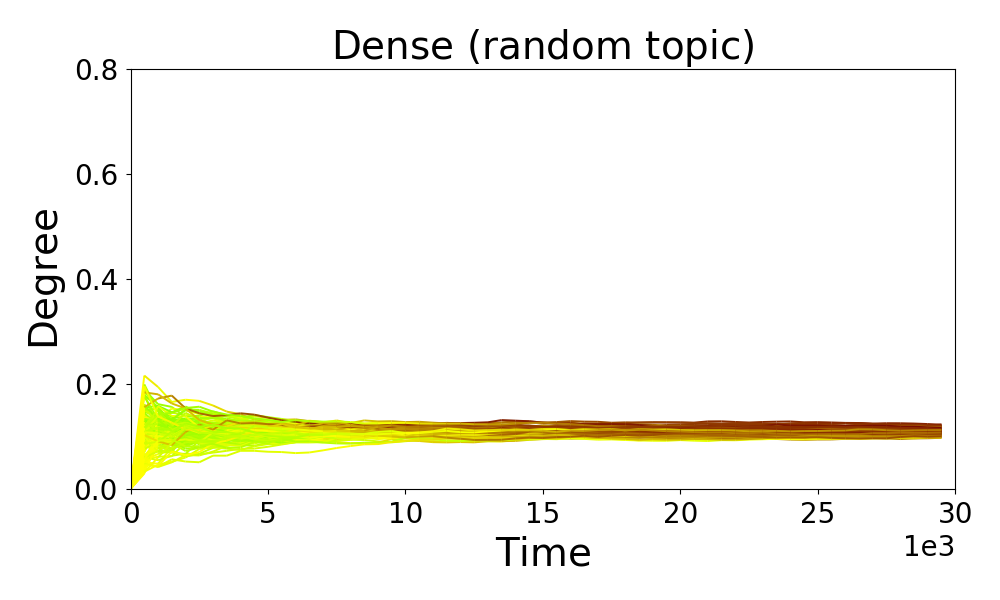}

\caption{Same as Figure \ref{fig:degree-time-AcqC} but with randomized topic choice for all ordinary agents.}
\label{fig:degree-time-noAcqC}
\end{figure}

\paragraph{High Centrality Without Successful Propaganda} As in the previous condition, the MIA can achieve high centrality in small-world and ring networks, though only low centrality in dense networks. However, in this condition, high centrality is no longer associated with high reputation for the MIA, as seen in Figures~\ref{fig:degree-rep-noAcqC} and~\ref{fig:degree-time-noAcqC}, which track the MIA's centrality alongside its eventual reputation and reputation over time, respectively. If the randomized topic choice had merely made the MIA irrelevant, its degree would have collapsed---but it does not: its centrality remains in the same range as before, and the correlations in Figure~\ref{fig:degree-rep-noAcqC} are still strong. Centrality therefore remains a key structural factor; it simply no longer translates into high reputation.


The reason is that ordinary agents now spend much more of their interactions on topics other than the MIA, allowing them to build knowledge and trust among themselves, counteracting the effect of echo-chambers. As the time evolution in Figure~\ref{fig:degree-time-noAcqC} shows, this pattern is highly robust and persistent among all network types: early on, the MIA might still appear honest\footnote{This is not surprising: uninformed, weak opinions at the beginning result in vague statements that naturally sound plausible. As they do not contradict with anything, there is no reason to mistrust.}, but it cannot sustain this for long.

\begin{figure}[H]
\centering
\includegraphics[width=0.32\textwidth]{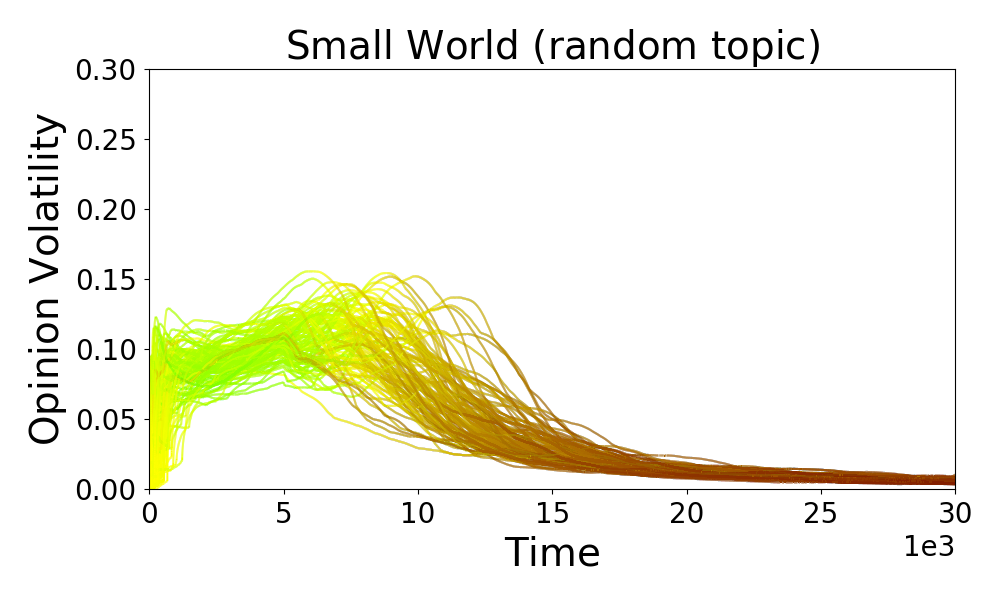}
\includegraphics[width=0.32\textwidth]{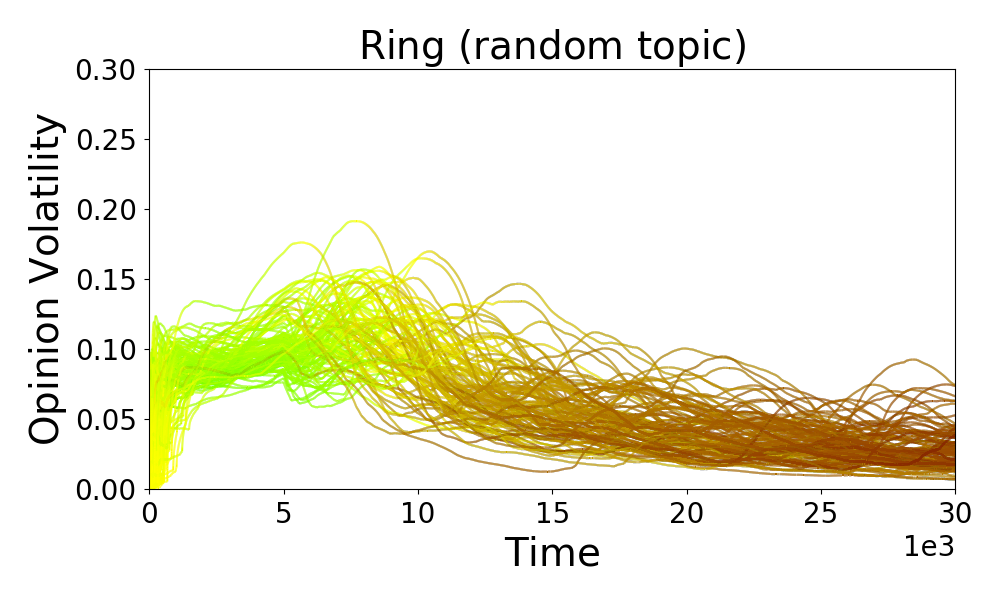}
\includegraphics[width=0.32\textwidth]{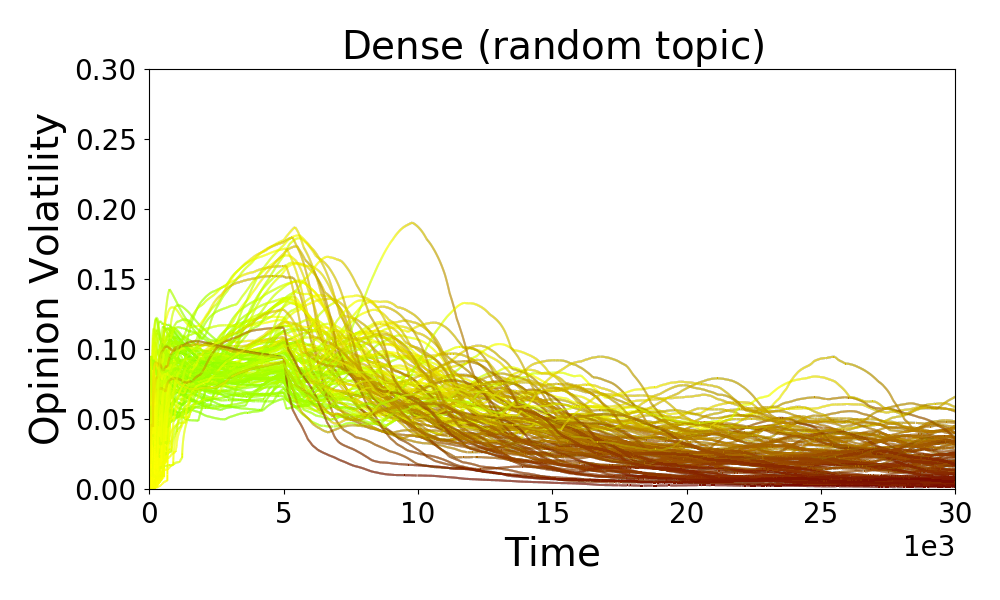}

\caption{Same as Figure \ref{fig:opinion-volatility-AcqC} but with randomized topic choice for all ordinary agents.}
\label{fig:opinion-volatility-noAcqC}
\end{figure}

\paragraph{Stability \& Consensus}
As one might already expect from this robust behavior, opinion formation in these scenarios is very calm---rather slow but consistent. Figure~\ref{fig:opinion-volatility-noAcqC} confirms that, after a moderately volatile phase at the beginning\footnote{Again just a consequence of flat initial opinions: while a single bit of new information causes a significant change in opinion initially, this effect vanishes as soon as opinions get more pronounced.}, opinion volatility settles down to a very low value in all network types and stays there for the rest of the simulation. The region of high uncertainty and rapid opinion changes seen in Figure~\ref{fig:opinion-volatility-AcqC}, associated with high reputation of the MIA especially in small-world networks, has vanished completely. Here, once the ordinary agents have developed a low opinion of the MIA (which they consistently manage within the first third of the simulation), they barely change their minds: their worldview remains consistent with high consensus among the entire network. 

\begin{figure}[H]
\centering
\includegraphics[width=0.32\textwidth]{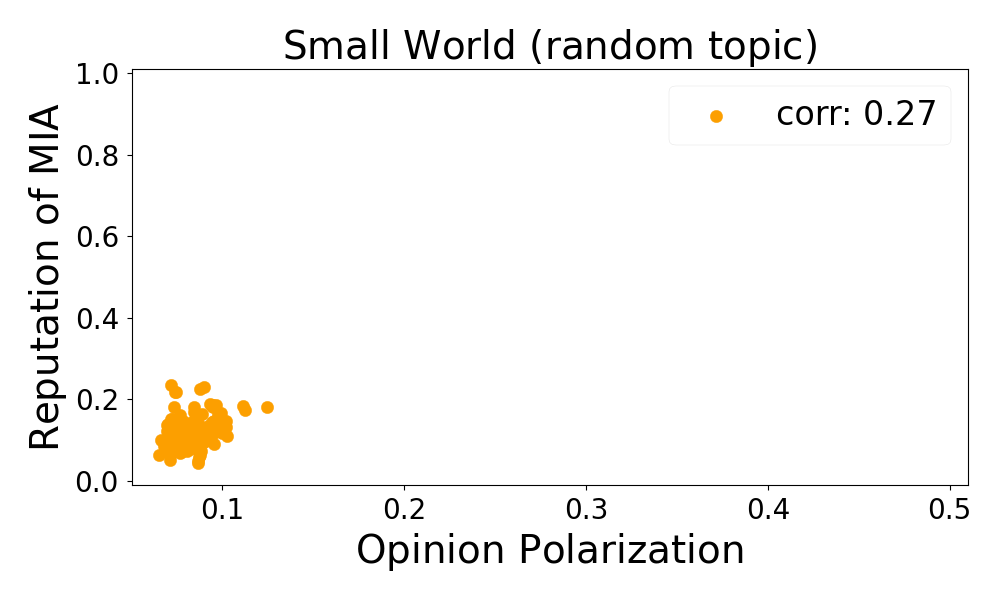}
\includegraphics[width=0.32\textwidth]{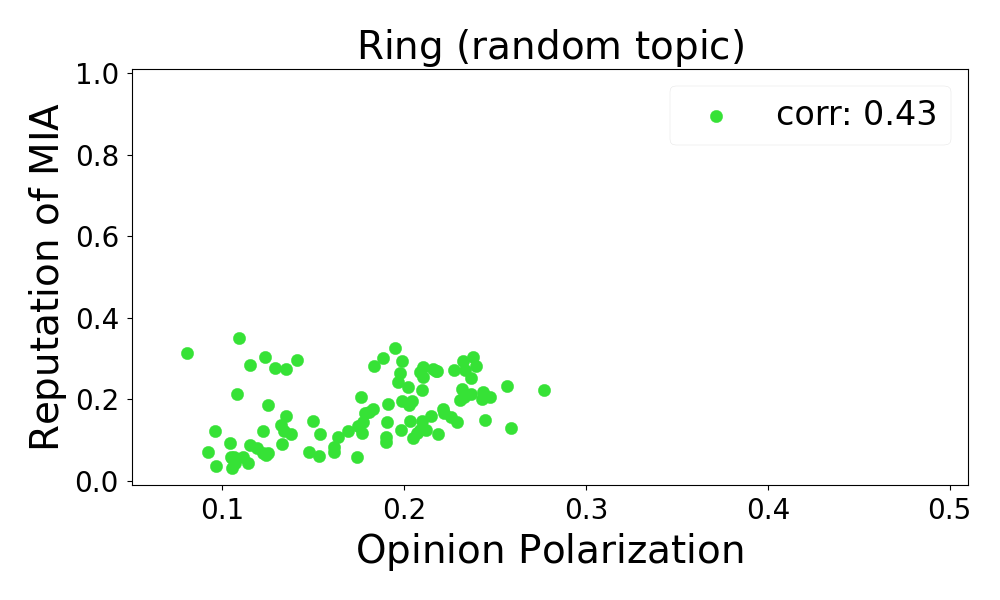}
\includegraphics[width=0.32\textwidth]{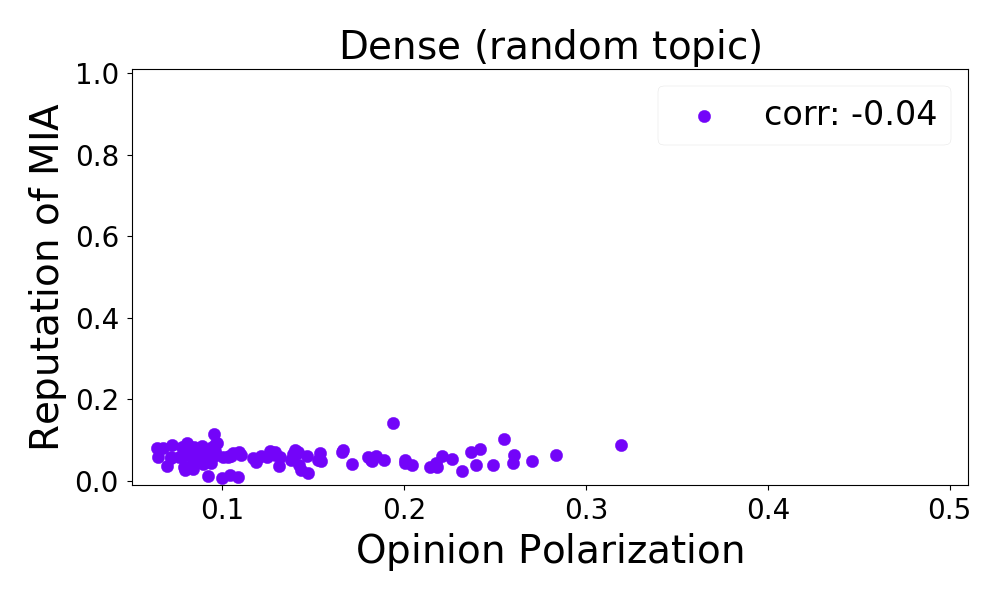}

\caption{Same as figure \ref{fig:polarization-AcqC} but with randomized topic choice for all ordinary agents.}
\label{fig:polarization-noAcqC}
\end{figure}

The final question is whether this process truly holds for all ordinary agents, or whether some still diverge. For this, we turn to the polarization data in Figure~\ref{fig:polarization-noAcqC}. In small-world networks, polarization stays very low, indicating shared consensus. In ring networks, polarization is slightly reduced compared to Condition~A but still shows some higher values and a moderate correlation with the MIA's reputation. Thus, the somewhat higher reputations around~$0.3$ (the wider spread for reputation in ring networks in Figure~\ref{fig:rep-hist-noAcqC}) stem from small subgroups rather than reflecting network-wide agreement---a natural consequence of the ring structure, where limited alternative interaction partners allow the MIA to retain disproportionate influence over a few individuals. 

Interestingly, in dense networks---despite their highly interconnected structure---polarization sometimes exceeds that in Condition~A. This, however, does not reflect disagreement about the MIA. As Figure~\ref{fig:rep-hist-noAcqC} shows, opinions about the MIA are highly aligned; otherwise reputations would not cluster so sharply at very low values. Instead, the polarization arises from divergent opinions about ordinary agents. Only in dense networks with randomized topics do individuals reliably meet almost everyone and actually discuss them, enabling meaningful opinions about many agents. These views, informed by different local experiences, do not necessarily align immediately, though they would likely converge over time---just as they already do for the MIA, about whom more information circulates.

\subsubsection{Interim Conclusion --- Condition B}

Compared to acquaintance-based topic choice, randomized topic choice provides a healthier, consensus-building and propaganda-resilient society. Both at the group and the individual level, agents encounter less friction and find themselves in a coherent rather than conflicting environment. The mere fact that agents talk about a broader range of topics (even if many statements only have little informational content), helps them calibrate opinions and become resilient against centralized propaganda, across all network structures. Importantly, this does not require ignorance or exclusion---the MIA still occupies a central role in the networks and its views are still shared widely. The agents' just distribute their attention instead of focusing on the MIA, which prevents them from amplifying and ultimately falling for its propaganda.

\subsection{Partial Randomization of Topic Selection and Individual Benefit}
\label{subsec: transition-from-A-to-B}

Having seen that distributing attention helps the network to resist propaganda, one important question remains: does this only work if all agents randomize their topic choices, or can a few individuals already make a change? The former requires a substantial systemic shift, whereas the latter would allow a group's epistemic standing to improve with more modest small-scale changes in behavior. 

\begin{figure}[H]
\centering
\includegraphics[width=0.32\textwidth]{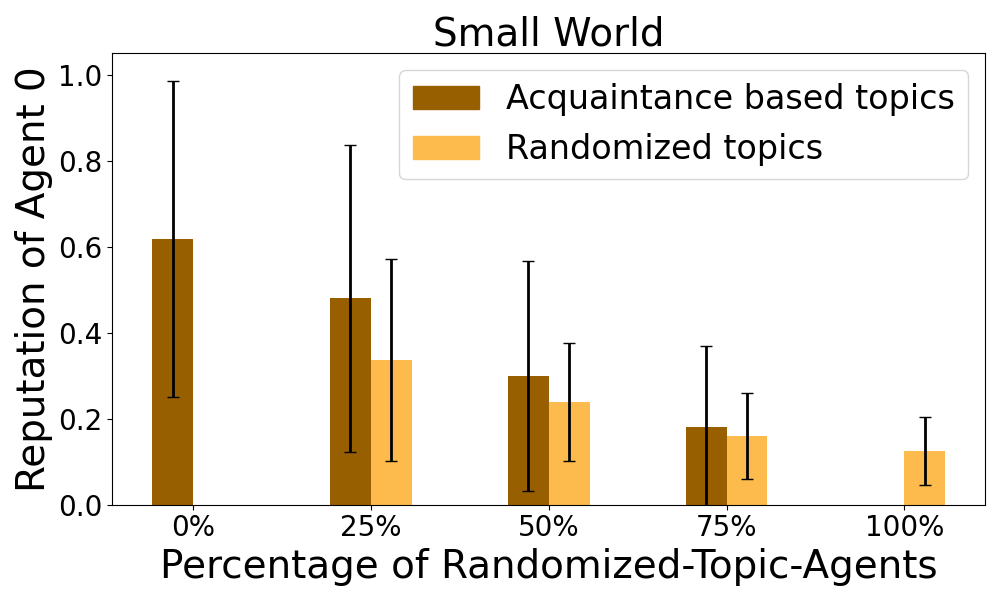}
\includegraphics[width=0.32\textwidth]{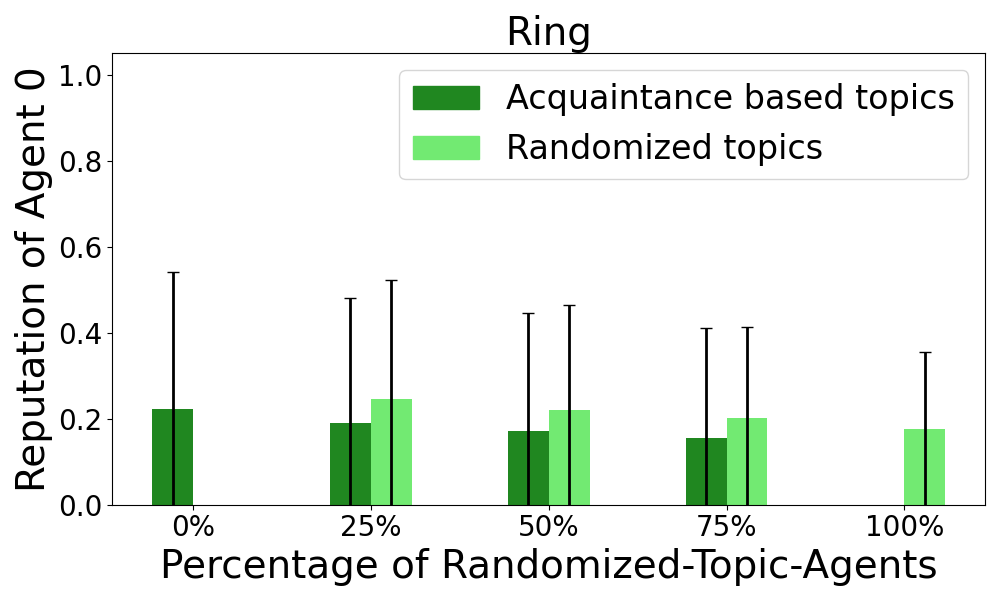}
\includegraphics[width=0.32\textwidth]{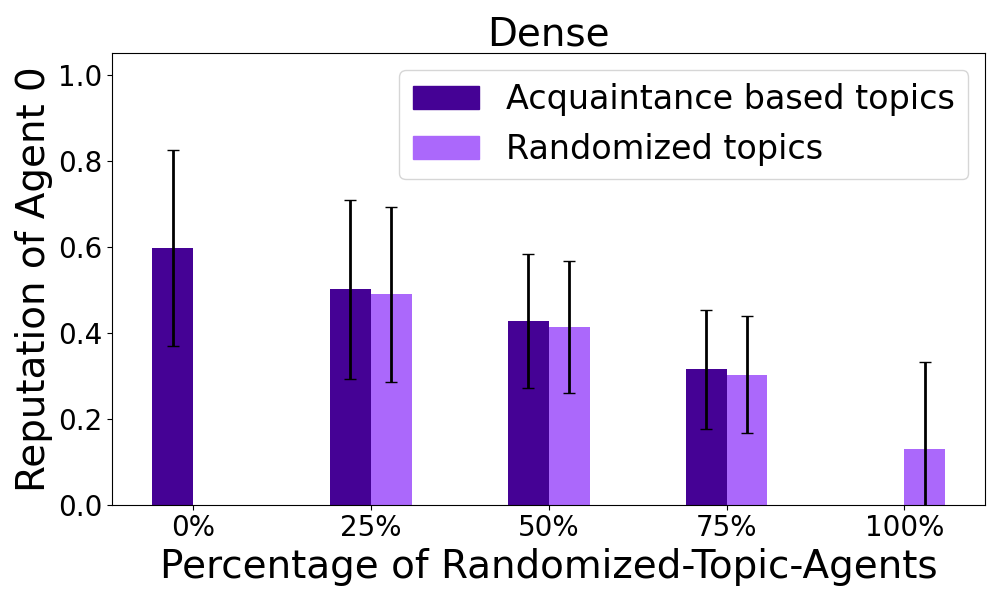}
\caption{Reputation of the MIA (agent $0$) under a varying percentage of ordinary agents that choose conversation topics based on acquaintance (darker bars) or at random (light). Different network types are reflected in each sub-panel. For each percentage and network type, results show the average of the individual ordinary agents' opinions at the end of $100$ simulations with different random seeds. Error bars indicate the standard deviation.}
\label{fig:percAcqC_rep0}
\end{figure}

Figure \ref{fig:percAcqC_rep0} shows that in small-world networks, even when a minority ($25\%$) of agents randomize their topics, this reduces the average reputation of the MIA. The opinions of randomized-topic ordinary agents (lighter bars) are closer to the ground truth than those of ordinary agents choosing based on acquaintance (darker bars), so this strategy benefits its users the most. Still, even those who do not adopt the randomized-topic strategy see an improvement when others---even a minority---do so. The network as a whole grows more resilient, and resilience increases as a higher proportion of agents choose randomly.

For dense networks we see a similar pattern: random allocation of attention via topic choice increases network resilience, even when only a minority do so. However, the overall effect is weaker, as is the difference between randomized-topic and acquaintance-based-topic strategies, so randomized-topic agents do not benefit as much from their own strategy as they do in small-world networks. As dense networks offer more pathways for information flow, each individual agent has less influence on overall dynamics, and randomized-topic agents see less epistemic benefit from their strategy. 

In ring networks, the MIA's reputation is consistently low, independently of the percentage of randomized-topic agents. The generally high variance here reflects the bimodal distribution previously seen in ring networks (Figure \ref{fig:rep-hist-AcqC}). Nevertheless, a slight reduction in variance is visible as the percentage of randomized-topic agents increases, with the extreme case ($100\%$ randomized-topic agents) reflecting the absence of bimodality seen in Figure \ref{fig:rep-hist-noAcqC}.

\section{Discussion}

Our model operationalizes limited attention as a constraint on the number and diversity of conversation partners and topics, as well as on memory, reflecting core cognitive principles such as bounded rationality and selective exposure. It shows how this limited cognitive resource impacts which information spreads and persists at the group level, similar to what other studies---both theoretical and empirical---have shown \cite{kahneman1973attention, friedkin2011social, bakshy2015exposure, qiu2017limited}. Thus, micro-level attention patterns shape emergent collective phenomena such as consensus or resilience to misinformation. 

Our results demonstrate that the susceptibility of a social system to misinformation is shaped by the emergent network structure, which---in our case---itself arises from the cumulative actions and preferences of individual agents, becoming a self-conserving, stable structure. In general, across various network structures, we find that an overall well informed population emerges in situations with high topic variability, but with low individual opinion volatility and polarization. This reflects a healthy informational environment where consensus is reached through stable and aligned individual opinions. 

Crucially, this alignment is not caused by any external pressure\footnote{Apart from a general small bias that considers others with similar opinions more trustworthy and thus rewards aligned opinions with less social conflict (as a high surprise leads to lower credibility of a message, see equation \ref{eq:credibility}).}. Nor is it caused solely by the existence of a central propagandist, as seen in the contrast between our two experimental conditions. 

In contrast, misinformed states have many possible causes depending on the properties of different network structures:

\begin{itemize}
\item In dense networks, low accuracy can arise even when the group appears unified and stable: when both volatility and polarization are low, the system may settle into a persistent but incorrect consensus, recalling phenomena such as groupthink and the false consensus effect \cite{janis1972victims, ross1977false}. In such a system, (mis-)information can spread slowly but steadily and mostly unnoticed.

\item Alternatively in ring networks%\footnote{Or, more generally, in networks that show long, unbranched connections, as this is the decisive feature of ring networks impacting dynamics}
, the group fragments into competing subgroups with unstable beliefs, showing high individual opinion volatility as well as high group-polarization. This combination of social fragmentation and substantive disagreement have been explored by other studies \cite{lin2023effects, minson2022exposure}, where usually the fragmentation is caused by disagreement. Our simulations show that the other direction is also possible: A given fragmentation of the network can lead to persistent disagreement, combined with low accuracy and incorrect opinions (at least for a significant part of the group). 

\item Low accuracy can also emerge in highly volatile environments with low polarization. When individuals frequently shift their opinions, this may lead to rapid and collective adoption of misinformation, as seen in informational cascades and herding behavior \cite{WANG2018441, bikhchandani1992theory, banerjee1992simple}. This was especially prevalent here in small-world network structures. As connectivity within the network is still high enough to bring everyone into the opinion formation process, these informational cascades usually come with low polarization. Importantly, the flexibility to focus a whole network's attention onto a single player enables rapid and widespread propagation of misinformation, illustrating how micro-level choices can scale up to macro-level vulnerabilities. As small-world networks are very common, and as transient high consensus (i.e., low polarization) makes it hard for individuals to spot potential misinformation, this risk should be taken seriously. Even if individuals manage to spot misinformation, the apparently widespread consensus around them (even if it is only supported by very volatile agents themselves) may make it hard to uphold this deviating opinion. 

\item Compared to small-world networks, dense and ring networks bring a small degree of natural resistance against strategic misinformation campaigns by not allowing for such extreme attention concentration.

\end{itemize}

Despite the diversity of these mechanisms driving misinformation, results showed that randomizing the topic of conversation can mitigate them all. By reducing the maximal attention that a single agent can get, it leads to higher accuracy across all network types. Diversifying one's conversation topics is thus a straight-forward rule that fosters resilience in every situation considered here. 

Although this might not be so easy for everyone to implement in their their every-day lives, it is still a simple rule---much simpler than trying to robustly identify a small number of trustworthy sources of information to rely on. Further, it does not depend wholly on having platform providers change their algorithms or implement content moderation. A further feature in its favor is that non-compliance with the rule is easy to detect: it is easier to notice that one has been talking about the same topic all day, than to track just how reliable all the information from one's conversation partners has been during those discussions. 

Moreover, it is not required that a whole population diversifies their topics to see any effect: especially in small-world networks, a small fraction of actors doing so already makes a difference. By paying off for the very individual that sticks to the rule, it appears to be an appealing approach with immediate rewards and a low barrier to entry.

These results may suggest hypotheses for experimental psychology, including interventions that prompt individuals to diversify their conversation topics in order to reduce susceptibility to misinformation and increase opinion stability and healthy consensus building. However, previous studies in similar directions have already found that such measures are highly content- and situation dependent and might even backfire if not handled carefully \cite{santoro2022promise, bail2018exposure}. Particular caution is required when topics are morally charged or social identities are challenged, all of which are currently not included in the Reputation Game Simulation and are therefore not reflected in the results. 

Apart from this topic-diversifying strategy, the model reveals a noteworthy interplay between individual and group levels. While individuals may frequently change their minds or experience uncertainty, the group can nonetheless arrive at a stable (though not necessarily accurate) consensus, highlighting the distinction between individual volatility and collective stability, especially in small-world networks. Seen from the inside, an apparently stable group opinion can seem powerful and compelling, although the single individuals (if they only knew) are not alone in their doubts about it. This aligns with empirical social science referring to social pressure or pluralistic ignorance where  people privately have different opinions to what they think the majority has, but still publicly process those majority opinions \cite{asch1955opinions, prentice1993pluralistic}.

Direct benefits could also be derived from the RGS if it was used as a test environment to investigate the effects of individual cognitive biases, such as confirmation bias or motivated reasoning, in the context of group dynamics. In complex networks, group-level outcomes are typically not straightforwardly and transparently inferrable from individual traits, and group outcomes at one time point may impact individual cognitive traits at a later time point. By allowing for systematic manipulation of agents' information selection or updating strategies, in combination with varying network structures (including allowing diverse structures to emerge via interactions, as here), the RGS enables controlled exploration of how and when individual cognitive biases translate into collective phenomena such as polarization, echo chambers, or resilience to misinformation. In particular, it allows researchers to disentangle the contributions of individual-level mechanisms from emergent network effects, thereby providing a clear bridge between cognitive models of reasoning and large-scale social dynamics.

\subsection*{Limitations \& Future Directions}
The Reputation Game Simulation is limited by a number of factors that make it a simplified and only partially transferable model of real-life societies. 

The current model had just one malicious agent pulling attention (though this allowed us to identify causal effects of this strategy). In reality,  numerous malicious information sources, both coordinated and competing,  draw attention \cite{jacobs2024whatisdemocracy, luceri2019red, marigliano2024analyzing}. The effect of coalitions of influence and inconsistent or conflicting misinformation campaigns are thus left for future work.

Decisions about whom to talk with, what to say and even how to interpret information coming from those various sources were also simplified, only capture acquaintance and reputation, apart from random choices. In human cognition, however, these choices reflect various motivations, social heuristics, and affective states \cite{Tversky&Kahneman, lerner2015emotion, perlovsky2010curiosity, mcpherson2001birds}. Introducing more psychologically grounded decision rules (e.g., based on homophily, trust, curiosity, or emotional salience) would connect the model to established theories of selective exposure, motivated reasoning, and affective decision-making.

Further, all agents' traits (apart from opinions and network connections) were fixed. Realistically, individuals change over time. Such changes may also reflect simultaneously evolving opinion dynamics. Exploring this would require more work at the intersection of social learning and cognitive adaptation. 
 
Agents currently rely solely on dyadic gossip. Human communication, in contrast, involves diverse formats including group discussions, institutions, and media systems, all with their own style of communication, intentions and range of topics. Incorporating multiple information channels and different (kinds of) topics would make the model more comparable to empirical studies of media influence, institutional trust, and multi-source integration, providing a computational framework to test how different communication modes, modalities and media interact to shape belief networks.

Finally, future work should examine how interventions that broaden topic selection or diversify social contacts---in general, decentralize attention---can be implemented in more realistic systems, considering the aforementioned challenges. Such simulations could test cognitive theories on attentional breadth and exploration/exploitation trade-offs, clarifying how encouraging informational diversity strengthens collective reasoning and resilience to misinformation.

\section{Conclusions}

The structure of communication networks, as shaped by individual behaviors, plays a decisive role in determining a society's vulnerability or resilience to misinformation. Concentrated attention and strategic communication can enable malicious, dishonest agents to gain influence, particularly in networks with small-world characteristics, which allow for heavy feedback cycles. 

Instead, bringing diversity into conversation topics---thereby distributing attention more evenly---has been shown to be an effective mechanism to counteract manipulative influence and support resilience. Although these insights arise from an abstract simulation and should be viewed as conceptual mechanisms rather than direct empirical predictions, they align with theories of bounded rationality and collective cognition, showing how local attention constraints scale to societal belief patterns.

In the end, results suggest a clear, actionable and robust principle that strengthens resilience against misinformation: distribute attention and diversify topics. 
\paragraph{Acknowledgements}
We gratefully acknowledge Céline Bœhm for her invaluable support and major contributions during the early stages of the model design, for many fruitful discussions throughout the project, and for carefully proofreading the manuscript.

\newpage
\bibliographystyle{unsrtnat}
\bibliography{references}

\newpage
\appendix
\section{Effects of the MIA's Individual Components}\label{sec: parameter-study}
As stated in \S \ref{subsec:experiment-specific-setup}, the MIA's strategy consists of three components that distinguish it from ordinary agents: its communication strategy (how to choose conversation partners and topics and what to say in each situation), switched off acquaintance preference for these choices and a reduced shyness. Here, we want to show each component's individual effect on agent 0's ability to influence all others' opinions.\\
For each component there are several options to compare: Acquaintance (by default switched on for both the choice of conversation partner and topic) is switched off, effectively flattening equations \ref{eq:draw_b} and \ref{eq:draw_c}. Shyness (by default at $S=10^{0.5}$) is lowered to $S=10^{0.3}$, increasing the agent's audience in one-to-many interactions to approximately $5$ or $6$ and decreasing acquaintance affinity\ --\ if switched on \ --\  according to equations \ref{eq:draw_b} and \ref{eq:draw_c}. The communication strategy (by default ordinary as described in \S \ref{subsec:the-reputation-game}) can be changed to the mass-influencing strategy described in \S \ref{subsec:experiment-specific-setup}, but since this is itself a combination of the ``manipulative'' and ``dominant'' strategies used in earlier studies \cite{ensslin22, ensslin2023simulating}, we will also consider these strategies individually. The manipulative strategy is what the MIA uses in one-to-one conversations: choosing dishonest conversation partners and flattering them, whereas the dominant strategy is what the MIA uses in one-to-many interactions: choosing honest recipients and talking positively about itself. When used individually, both strategies are applied no matter the interaction type.\\
A summary of all components and possible values is given in table \ref{tab: MIA-components-doe}, leading to $16$ configurations of agent 0 in total. As a network structure we use the small-world setup only, as it is the most common one \cite{de1978contacts, schnettler2009structured, milgram1967small, mislove2007measurement, dimitri2023facebook, kwak2010twitter} and also to avoid increasing the number of combinations too much.\\
\begin{table}[ht]
\centering
\renewcommand{\arraystretch}{1.3}
\begin{tabular}{l l}
\hline
\textbf{Component} & \textbf{Values} \\
\hline
Acquaintance & On \;\textbar\; Off \\
Shyness & Default: $S=10^{0.5}$ \;\textbar\; Low: $S=10^{0.3}$ \\
Communication Strategy & Ordinary \;\textbar\; Dominant \;\textbar\; Manipulative \;\textbar\; Mass-Influencing \\
\hline
\end{tabular}
\caption{Components of the MIA's strategy with possible values to compare against ordinary agents.\label{tab: MIA-components-doe}}
\end{table}

\begin{figure}[H]
\centering
\includegraphics[width=0.7\textwidth]{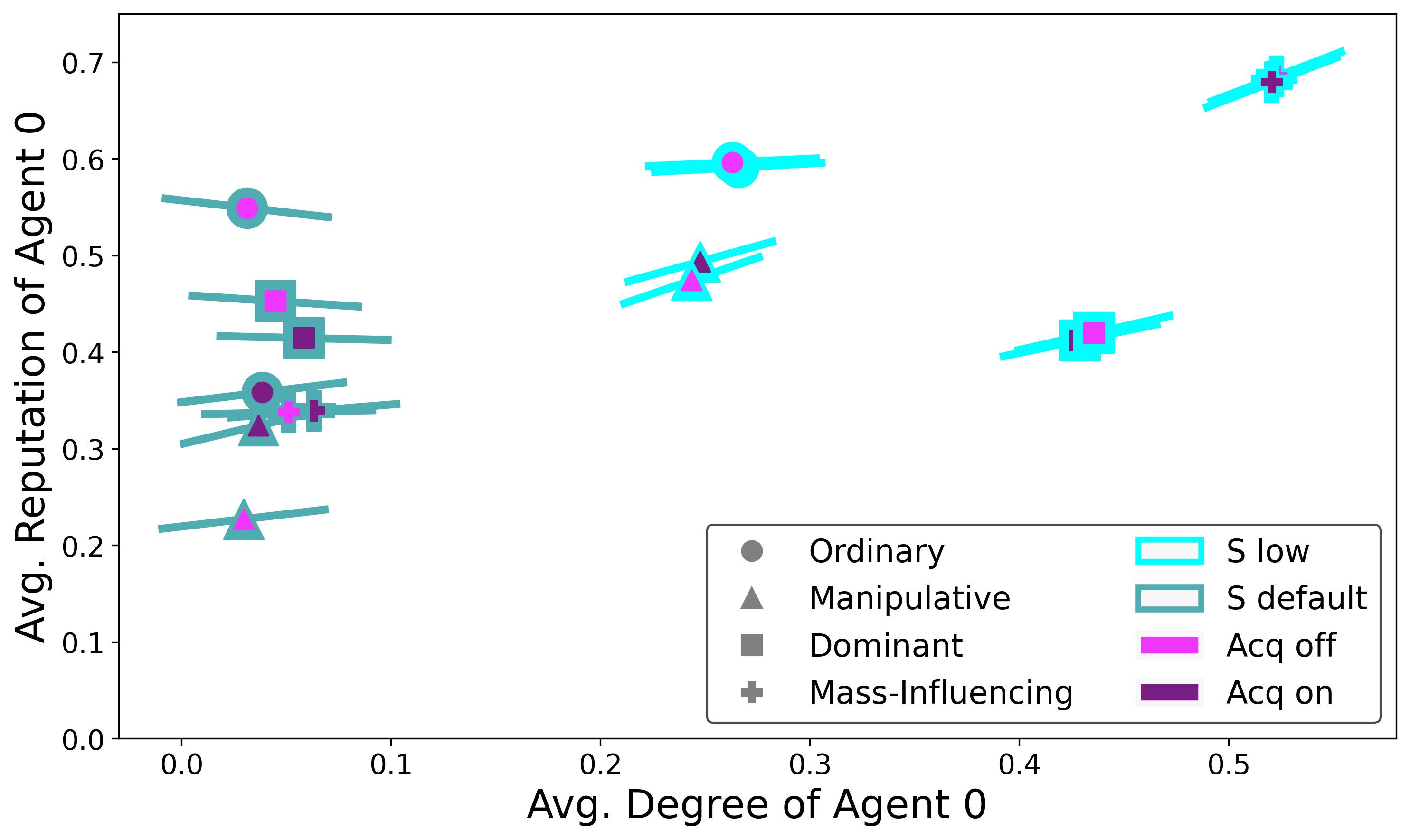}

\caption{Average degree of agent 0 compared to its average reputation for varying configurations shown in table \ref{tab: MIA-components-doe}. Averages are taken over the final states of $100$ simulations with different random seeds for each configuration. The shape indicates agent 0's communication strategy, the outer color its shyness and the inner color whether or not it uses acquaintance for decisions. All other agents are ordinary and configured to form small-world-like network structures. The bars behind each point indicate the correlation between agent 0's degree and reputation for the specific configuration in the depicted units. A high slope means high correlation, negative slope means negative correlation and a horizontal bar means no correlation.}
\label{fig: appendix-degree-rep}
\end{figure}

\begin{figure}[H]
\centering
\begin{subfigure}{0.32\textwidth}
  \includegraphics[width=\linewidth, trim=80 50 80 50,
    clip]{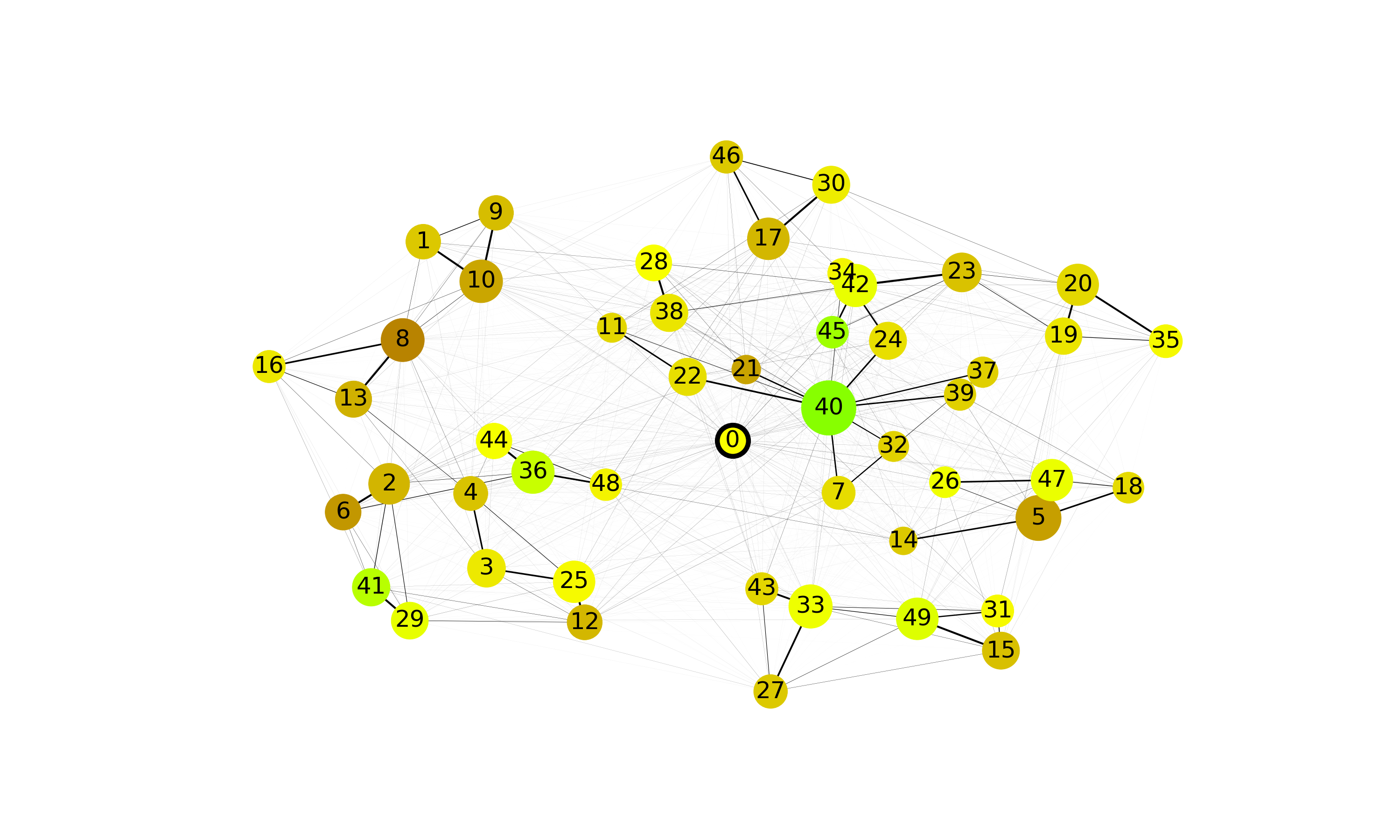}
  \caption{\centering S default, Acq off}
\end{subfigure}
\begin{subfigure}{0.32\textwidth}
  \includegraphics[width=\linewidth, trim=80 50 80 50,
    clip]{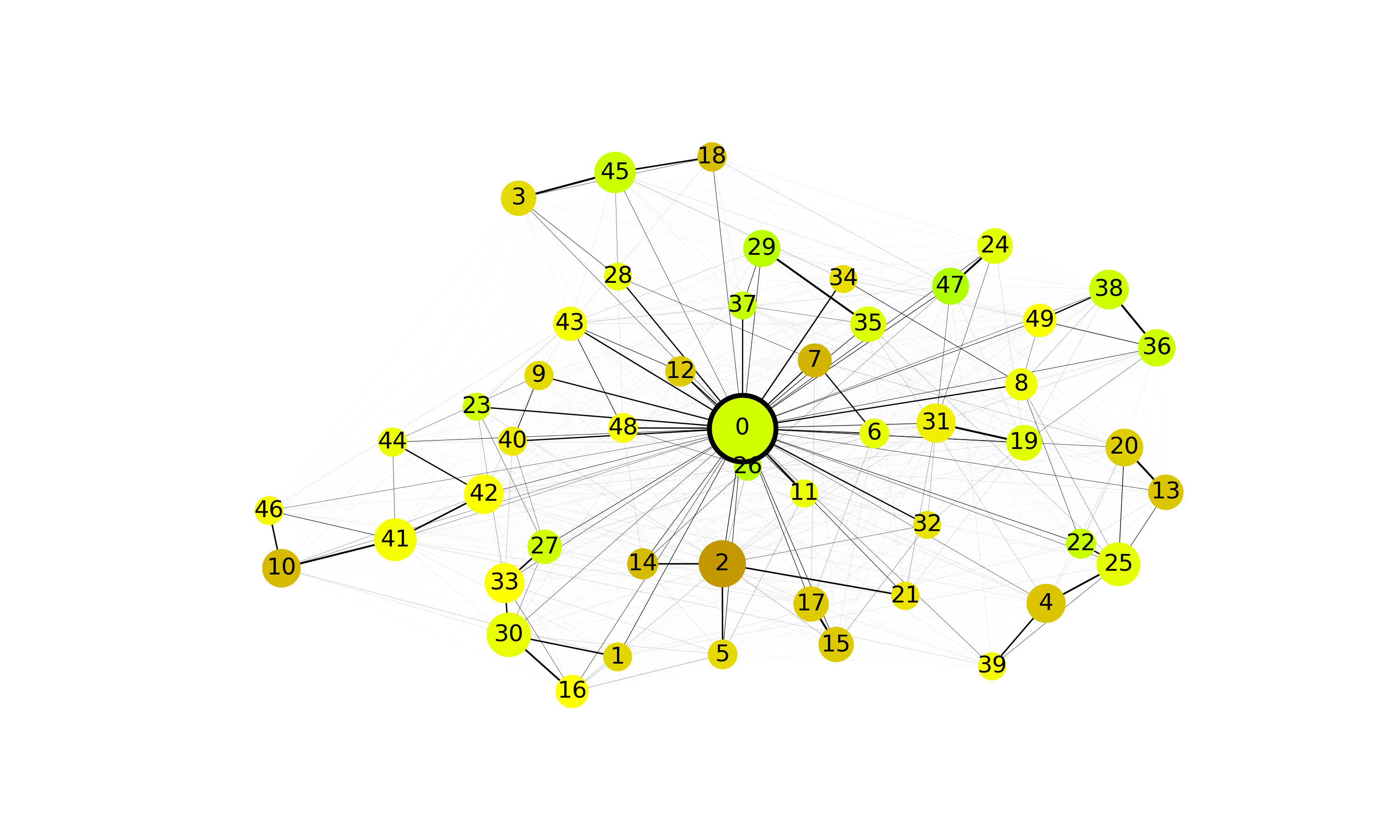}
  \caption{\centering S low, Acq on}
\end{subfigure}
\begin{subfigure}{0.32\textwidth}
  \includegraphics[width=\linewidth, trim=80 50 80 50,
    clip]{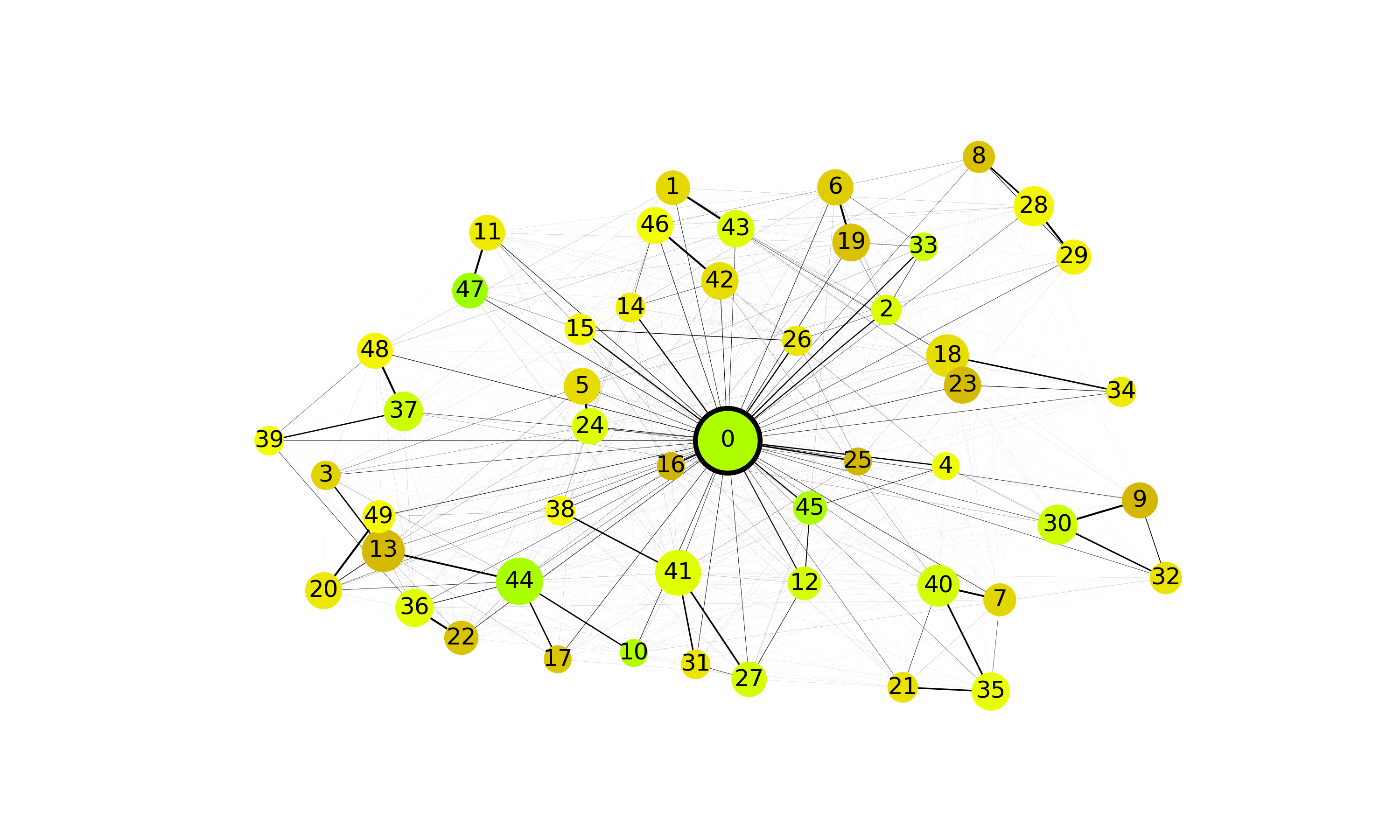}
  \caption{\centering S low, Acq off}
\end{subfigure}

\caption{Exemplary network structures for three configurations deviating from default values as specified by the sub-captions (the network structure with default values only has already been shown in Figure \ref{fig:stability-against-mia-network-types}d). The communication strategy used by agent 0 is ordinary in all cases, but since shyness and acquaintance are the main structure-giving parameters, the networks look similar for other strategies.\label{fig:appendix-networks}}
\end{figure}

For the evaluation we consider the final reputation of agent 0\ --\ measuring how successfully it convinced others of its honesty despite being fully deceptive\ --\ as well as the emerging network structure, quantified by agent 0’s degree at the end of the simulations. Both measures are shown and put into relation in Figure~\ref{fig: appendix-degree-rep}, while Figure \ref{fig:appendix-networks} illustrates exemplary network structures that emerge.\\
Across all strategies, Figure \ref{fig: appendix-degree-rep} shows that low shyness is the key factor for achieving a high degree and typically also a higher reputation. This is, because a low shyness lets agent 0 sustain many (strong) contacts and directly influence large parts of the network as shown in Figures \ref{fig:appendix-networks}b,c for the ordinary strategy as an example. However, the effectiveness of low shyness in terms of reputation strongly depends on the communication strategy. The manipulative and dominant strategies perform worse than the ordinary strategy in this setting, indicating that their strategic communication is even disadvantageous: For the manipulative strategy, flattering messages\ --\ now also used in one-to-many interactions\ --\ can only be tailored to a single receiver and therefore easily appear inconsistent or implausible to others, facilitating lie detection. The dominant strategy, as it frequently makes itself the topic of conversation and thus causes others to disproportionately choose the dominant agent themselves as partner and topic, yields a pretty high degree, but this does not seem to automatically help with reputation. Instead, the dominant agent has little options to gather information about others, thus can only place lies poorly, and in turn is caught more often, limiting its propagandistic success. Combining both approaches in the mass-influencing strategy, however, overcomes these weaknesses, outperforming all other configurations in both reputation and degree.

Some strategies (notably ordinary and dominant) also achieve moderately high reputations at default shyness. However, their impact on the whole system remains limited, which becomes evident when looking at emerging network structures. In case of switched off acquaintance (Figure \ref{fig:appendix-networks}a), agent $0$ forms many, but only weak ties, resulting in flat opinions close to the initial mean of $0.5$ for most agents. In case of switched on acquaintance (already seen earlier in Figure \ref{fig:stability-against-mia-network-types}d), it maintains strong, but only a few connections within a small subnetwork, so high reputation is driven by only a small number of agents. In either case, no large network effects occur.

\section{Evaluation Metrics}\label{sec:metrics}
The metrics used for quantifying the dynamics are reputation, degree, and opinion volatility of individual agents, as well as opinion polarization on group level, which are defined in the following.

\paragraph{Reputation} is the perceived honesty of an agent by the others. The higher an agent's reputation, the higher the credibility of its messages and the higher its influence on others' opinions. In order to calculate the reputation of an agent $i$ in the eyes of all others, we calculate its expected honesty according to the overall group opinion about that agent $P_\mathrm{group}$
\begin{align}
\mathrm{reputation}_i &= \langle x_i \rangle_{P_\mathrm{group}(x_i)}.
\end{align}
Assuming that all agents' opinions about the agent $i$ are independent from each other, the group opinion
\begin{align}
P_\mathrm{group}(x_i) &\propto \prod_{j \in \mathcal{A} \setminus \{i\}} P(x_i|I_{ji}) \propto \prod_{j \in \mathcal{A} \setminus \{i\}} x_i^{\mu_{ji}}(1-x_i)^{\lambda_{ji}}
\end{align}
reflects the differently strong opinions the agents have. Note, that for example totally uninformed agents that have a flat opinion distribution do not influence the group opinion at all, since their contributions are multiplications by $1$. With this, we can now calculate the expected honesty of agent $j$ and find 
\begin{align}
\label{eq:reputation_group}
\mathrm{reputation}_i = \frac{\mathcal{B}(2+\sum_{j \in \mathcal{A} \setminus \{i\}} \mu_{ji}, 1+\sum_{j \in \mathcal{A} \setminus \{i\}} \lambda_{ij})}{\mathcal{B}(1+\sum_{j \in \mathcal{A} \setminus \{i\}} \mu_{ji}, 1+\sum_{j \in \mathcal{A} \setminus \{i\}} \lambda_{ji})},
\end{align}
with $\mathcal{B}$ being the beta function and the denominator resulting from normalization. By construction, this measure reflects the opinions of all agents about $i$ weighted with their individual certainty.

\paragraph{Degree} expresses the connectivity of agents. 
It is measured considering the total network structure, where each agent resembles a node and each pair of agents that talked to each other at least once is connected by an edge. The connection strength of these edges is then given by their number of conversations, where, for simplicity, we average the number of incoming and outgoing conversations between each pair of agents, as this might not always be symmetric due to one-to-many conversations. Further, since the number of conversations is growing steadily throughout the simulation, the overall connection strengths between all pairs of agents would become larger over time. While this might be a very legitimate way of description, we aim for a measure that reflects more the topology or overall shape of the network and less the total connection strengths. Therefore we normalize by the total number of conversations that happened up to a certain point in time $n^* = \sum_{i< j}n^\mathrm{c}_{ij}$. With this, we can measure the centrality of an agent $i$ in the network as its degree, given by
\begin{align}
\label{eq:degree}
\mathrm{Degree}_i = \frac{1}{n^*}\sum_{j \in \mathcal{A} \setminus \{i\}}(n^\mathrm{c}_{ij}+n^\mathrm{c}_{ji}).
\end{align}
This can be viewed as a weighted version of the frequently used degree centrality, here, basically measuring the fraction of all conversations agent $i$ took part in\ --\ in other words, how much it dominates the network conversation-wise.\\
Other frequently used centrality measures such as closeness or betweenness centrality would all reflect the same characteristics of the network structure, since the by far dominating network effect of the MIA is to form a star-like network around itself (cf.\ Figure~\ref{fig:stability-against-mia-network-types} in \S\ref{subsec:emergent-network-structures}). For our purpose, it is thus sufficient to only consider the agents' degree.

\paragraph{Opinion Volatility} describes how much the opinion of an agent changes over time in a certain interval. Specifically, we calculate the standard deviation of an opinion $\overline{x}_{ij}(t)$ within a certain time-window $T$
\begin{align}
\label{eq:volatility}
\mathrm{Opinion\ Volatility}_{i,j,T} = \sqrt{\left\langle \left( \overline{x}_{ij}(t) - \left\langle \overline{x}_{ij}(t) \right\rangle_T \right)^2 \right\rangle_T},
\end{align}
which measures how much the opinion of agent $i$ about agent $j$'s honesty changes or even fluctuates during this time interval. In this study we use an interval of $T=5000$ conversations.

\paragraph{Opinion Polarization} measures how diverse opinions among a group of agents are. Polarization is low if all opinions roughly align and high if there are contradicting opinions. Let us define $\mathbf{x}_i = (\overline{x}_{i1}, \overline{x}_{i2}, ..., \overline{x}_{iN})$ be the vector of agent $i$'s estimations of all other agents' honesties, i.e. agent $i$'s opinion. Then we can calculate the average opinion of the whole system by averaging over all agents' opinions
\begin{align}
m = \frac{1}{|\mathcal{A}|}\sum_{i \in \mathcal{A}} \mathbf{x}_i.
\end{align}
Further, the covariance matrix
\begin{align}
D = \frac{1}{|\mathcal{A}|}\sum_{i \in \mathcal{A}} (\mathbf{x}_i - m)(\mathbf{x}_i - m)^T 
\end{align}
represents the shape, orientation, and spread of the opinion distribution. If all opinions roughly align, we get a small sphere around that common opinion in an $|\mathcal{A}|$-dimensional space. If all opinions point into different directions but are homogeneously distributed we get a larger, but still approximately round sphere. If, however, opinions point into opposite directions, say especially for a few specific topics, the distribution will elongate along those directions and will become an high-dimensional ellipsoid. The more elongated the ellipsoid (i.e., the more extended into a few directions), the higher the polarization. In order to quantify the polarization this way, we calculate the ratio between the largest eigenvalue (a measure for how elongated the ellipsoid is in this direction) and the average eigenvalue (a measure for the overall spread or size of the ellipsoid across all directions)
\begin{align}
\label{eq:polarization}
\mathrm{Polarization} = \frac{|\mathcal{A}|}{|\mathcal{A}|-1} \frac{\mathrm{max}((d_k)_k) - \overline{d}}{\sum_k d_k},
\end{align}
where $k$ counts the dimensions of the covariance matrix, with $d_k$ being the eigenvalue along dimension $k$ and $\overline{d}$ is the mean of all eigenvalues. This way, polarization is highest when there is one (or a few) topics/agents in the system, about whom other agents have extremely diverging opinions.\\

\end{document}